\documentclass[12pt]{article}
\usepackage{epsf}
\input {./defs.inp}

\begin{document}

\begin{titlepage}

\centering

\protect\Huge\rm 1996/1997 LOW LUMINOSITY STATE AND STATE TRANSITIONS
OF GRS 1915+105: RXTE OBSERVATIONS
\vspace{17.mm}

\protect\large\rm S.~Trudolyubov$^{1,2}$, E.~Churazov$^{1,2}$,
M.~Gilfanov$^{1,2}$
\vspace{17.mm}

\large\it $^{1}$ Space Research Institute, Russian Academy of Sciences,
Moscow, Russia
\vspace{3.mm}

\large\it $^{2}$ Max-Planck-Institut f\"ur Astrophysik, Garching bei
Munchen, Germany
\vspace{7.mm}

\end{titlepage}

\clearpage

\section*{ABSTRACT}
The results of X-ray observations of Galactic superluminal jet
source GRS 1915+105 during its low luminosity state and state transitions 
in October, 1996 -- April, 1997 with {\it PCA} and {\it HEXTE} instruments 
aboard {\it Rossi X-ray Timing Explorer} ({\it RXTE}) are reported.
 
Except some peculiarities, the major spectral and temporal properties 
of the source during this period were similar to that of the Galactic 
black hole candidates in the {\it intermediate} state (corresponding to
the transition between the canonical {\it high} and {\it low} states).
The power law component of the spectrum gradually hardens with
decrease of the source flux in the $3 - 150$ keV energy band. The 
$2 - 10$ Hz QPOs were found to be a generic feature of the power
density spectrum in this state. {\it The changes of the QPO centroid
frequency are strongly correlated with the changes of spectral 
and timing parameters (in particular parameters of the soft 
spectral component). This type of correlation holds on the wide 
range of time scales (from $\sim$ seconds to $\sim$ months).} 
Overall trend of the spectral and temporal properties suggests 
that with further drop of the source flux (below the values observed 
during the reported period) they would match the canonical values 
for a {\it low} state of the Galactic black hole candidates.

\clearpage

\section*{Introduction}

The X-ray source GRS 1915+105, one of two Galactic objects which show
the superluminal expansion of matter, was discovered by {\it GRANAT} 
observatory as a transient in 1992 (Castro-Tirado, Brandt $\&$ 
Lund 1992). Radio observations of superluminal motion have allowed 
to determine the distance ($\sim
12.5$ kpc) and inclination of the system (Mirabel $\&$ Rodriguez
1994). The source is believed to be a black hole candidate on 
the basis of its observed high luminosity exceeding the Eddington 
limit (Sazonov \etal 1994) for a neutron star and similarities to 
another Galactic superluminal jet source GRO J1655-40 
(Zhang \etal 1994), whose dynamical mass estimate ($\sim 7 M_{\odot}$) 
implies the presence of a black hole (Orosz $\&$ Bailyn 1997). 
Long-term temporal behavior of GRS 1915+105 is very 
complicated: since the discovery a number of outbursts and low 
luminosity episodes have been reported (Sazonov \etal 1996; 
Paciesas \etal 1996).

Since April, 1996 GRS 1915+105 is a target of the {\it RXTE} TOO 
regular observations, which have revealed rich character of the source 
transient activity including the occasional complicated patterns 
of dips, flares and rapid transitions between high and low intensity 
alternated with the relatively quiet periods and also strong 
quasi-periodic oscillations (Greiner, Morgan $\&$ Remillard 1996; 
Chen \etal 1997; Belloni \etal 1997{\em b}; Morgan \etal 1997). 

We have performed the systematic study of the evolution of GRS
1915+105 properties using the data of PCA and HEXTE instruments 
aboard {\it RXTE} {\bf i)} to resolve the mutual dependence of the spectral 
and timing parameters of the source emission and {\bf ii)} to understand 
how the low luminosity state of GRS 1915+105 observed in October, 
1996 -- April, 1997 fits the standard scheme of the 'canonical' 
{\it high/low} states established for Galactic black hole candidates 
(Tanaka $\&$ Lewin 1995).

In this paper we report the results of the GRS 1915+105 {\it RXTE} 
observations during its low luminosity state and state transitions 
in October, 1996 -- April, 1997. The discussion of implications of
these data on to the physical models will be presented in the
subsequent paper (Trudolyubov \etal 1998).

\section*{Instruments and observations}
The observations discussed below were performed with the Proportional
Counter Array (PCA) and the High Energy Timing Experiment (HEXTE) aboard 
{\it Rossi X-ray Timing Explorer (RXTE)} (Bradt, Swank, $\&$ Rotshild
1993) in October, 1996 -- April, 1997. The list of observations is 
presented in Table 1. 

For the processing of the PCA data we used standard {\it RXTE} FTOOLS
version 4.1 tasks. For the background subtraction the version of the 
background estimator program taking into account the effects of activation 
due to the South Atlantic Anomaly, X-ray background and particle background 
based on the measurements of the Q6 rate (Stark 1997) was used. 

For the spectral analysis of the PCA data we used the 3.2.1 version of the
response matrix. In order to account for the uncertainties of the 
response matrix, a 1$\%$-- systematic error was added to the
statistical error for each PCA energy channel. We have
excluded from the spectral analysis the data below $3$ keV because 
of the rapid drop of the PCA effective area in this energy domain. 
In addition, because of the strong influence of the response 
uncertainties at high energies, the data above $20$ keV were also 
ignored. To evaluate the correct source flux, the standard dead--time 
correction procedure (Zhang $\&$ Jahoda 1996) was applied to the PCA data. 

HEXTE data also have been processed using the standard FTOOLS version 4.1 
tasks. We used latest available HEXTE response matrices, released on April 
3, 1997 and standard off--source observations for each cluster of
detectors to subtract background. In order to account for the
uncertainties in the response and background determination, only
data in the $20 - 150$ keV energy range were taken for the spectral 
analysis.

\section*{Results}
The {\it RXTE}/ASM light curve of the GRS 1915+105 in the $2 - 12$ keV
energy band is shown in figure 1. Prior to the end of October, 1996 (MJD
$\sim 50386$) the source was in the bright highly variable so called
'chaotic' state with average flux $\sim 1$ Crab (Morgan \etal 1997) 
and then has underwent transition to the low luminosity state (LLS) 
lasting $\sim 200$ days. The corresponding X-ray luminosity of the
source dropped from $\sim 10^{39} erg/s$ to $\sim 2 \times 10^{38} 
erg/s$ in the $3 - 20$ keV energy band (assuming the distance of 
12.5 kpc). After the April 25, 1997 (MJD 50564) GRS 1915+105 appeared 
to return to the high luminosity state (HLS). Contrary to the HLS 
with its high level of variability, the evolution of the source flux 
during LLS can be described as a relatively smooth decline to the 
$\sim 200 - 250$ mCrab level during first $\sim 100$ days followed 
by the similar--fashioned slow rise on the similar time scale. 

We present below the basic results of the systematic spectral and 
timing analysis of the source emission during the low luminosity 
state and state transitions using the data of PCA and HEXTE
observations (Table \ref{obslog}). 

\section*{Spectral analysis}
As we are interested in the general character of the GRS 1915+105 
spectral evolution, we used only simplest models (a sum of a 
multicolor disk black body model \footnote{Note that no corrections 
for the electron
scattering and effects of general relativity were made (Shakura $\&$ 
Sunyaev, 1973, Shimura $\&$ Takahara 1995), T is a measured color 
temperature} and a power law model with exponential cutoff corrected for
the interstellar absorption) to approximate its energy spectrum. 
Although this model provides satisfactory description to the overall 
shape of the spectrum, there is a significant excess of emission in
the $6 - 8$ keV region (with an average ratio of excess to the 
model continuum of $\sim 2-3 \%$), which is attributed to the presence
of the iron emission/absorption complex. Due to the relatively low 
energy resolution of the PCA instrument ($\Delta E \sim 1$ keV in the 
$6 - 8$ keV range) it is not possible to carry out a detailed analysis 
of this spectral feature. The presence of the iron emission/
absorption features shows the evidence for the additional
'reprocessed' component in the source spectrum, but its approximation 
by certain type of model requires further physical justification, so 
we have decided not to include 'reprocessed' component to the fit. 
\footnote{See Greiner \etal 1998 for the results of fitting of GRS
1915+105 LLS spectra with inclusion of the Compton reflection model} 
Because of the large uncertainty in the PCA response in the low 
energy domain, we fixed the equivalent hydrogen column density in the 
spectral fitting at the value of $\sim 5.0 \times 10^{22} cm^{-2}$, 
determined from ASCA observations (Ebisawa \etal, 1995). Due to the 
present difference in the relative normalizations of the PCA and HEXTE 
spectra, we used PCA normalization to compute broad-band spectral 
model fluxes.

We have generated the energy spectra of the source averaging the data 
over the whole observations. To trace the evolution of the spectral 
parameters during the individual observations with relatively high 
level of X-ray flux variability, additional analysis of the spectra, 
accumulated in the $16 - 80$ {\it s} time intervals, has been 
performed \footnote{Only PCA data were used for the analysis because
of the insufficient statistical significance of the HEXTE spectral
data accumulated during these time intervals}. This procedure has 
allowed us to study the evolution of the main spectral parameters of 
GRS 1915+105 in a wide range of time scales from {\it tens of 
seconds} to {\it months}.

For comparison with the low luminosity state (LLS) observations, we have 
analyzed the data for several observations of GRS 1915+105 covering 
the period of high luminosity state (HLS) (October 7,13,15 1996 observations) 
occurred just prior to the transition to the LLS. The typical 
broad-band spectra of the GRS 1915+105 in the HLS (in units of $F(E) 
\times E^{2}$) are shown in Figure \ref{spectra} ({\em left panel}). 
The source exhibits extremely complicated pattern of the spectral 
variability on the different time scales (Belloni \etal 1997{\em a}, 
Belloni \etal 1997{\em b}, Taam \etal 1997). The most striking 
feature of the spectral variation is the existence of two distinct 
types of broad-band spectrum corresponding to the highest ('flare') 
and lowest ('quescence') levels of source luminosity (Figure \ref{spectra}). 
The energy spectrum in the $3 - 150$ keV energy range can be well 
represented as a composition of the soft and hard components 
(Figure \ref{spectra}). The form of the hard spectral component 
in general is described by a power law (photon index $\alpha 
\sim 2.0 - 2.6$) which becomes steeper ($\alpha \sim 3.0 - 3.5$) 
at around $\sim 25 - 30$ keV. During the 'flare' the slope of 
the high energy part of the spectrum is $\sim 3.0 - 3.5$. The 
measured color temperature of the soft component is noted to be 
higher in the 'flare' state (see Belloni \etal 1997{\em b}).

Since October 23, 1996 GRS 1915+105 has begun a transition to the low
luminosity state lasting till November 28. During this time the 
source shows a rich character of the spectral variability 
alternating relatively quiet periods of low X-ray flux with a 
bright flaring episodes. As an example of the short-term spectral 
variability of GRS 1915+105 during the transition to the LLS the 
evolution of the basic spectral parameters of the source for the 
November 7, 1996 observation is shown in Figure \ref{pca_071196}. 
It is clearly seen that the changes of the X-ray flux are 
connected with the significant change in the slope of the high 
energy part of the spectrum and large variations of the soft 
component parameters. The hardness of the high energy part of 
the source spectrum expressed in terms of the best fit power 
law model parameters is closely related to the total X-ray 
flux (Figure \ref{pca_general}, \ref{hexte_general}; \ref{pca_071196}). In
addition, when the luminosity of the source is low enough, the 
cutoff of the spectrum becomes detectable with HEXTE at the 
energies of $\sim 70 - 120$ keV (Figure \ref{hexte_general}).

The broad-band $3 - 150$ keV energy spectrum of GRS 1915+105 in
the low luminosity state is also satisfactory described by the sum 
of two components: relatively weak soft thermal component with 
a characteristic temperature $kT \sim 1.0 - 1.7$ keV) and strong 
hard component which has approximately power law form ($\alpha 
\sim 1.8 - 2.4$) in the $20 - 60$ keV interval and cut-offs at 
around $\sim 70 - 120$ keV (Figure \ref{spectra}, {\em right panel}). 

The results of fitting of the HEXTE data with a power law model and 
PCA data with a composition of a power law and multicolor disk 
black body model for the observations covering the period from 
October 7, 1996 to April 25, 1997 (state transitions and low
luminosity state) presented in figures \ref{pca_general} and 
\ref{hexte_general} respectively. There is a strong correlation of 
the slope of high energy part of the spectrum and the source X-ray 
flux (Figure \ref{alpha_flux}).

\section*{Timing analysis}
To perform the analysis of the GRS 1915+105 timing properties 
we generated power density spectra in the $0.01 - 50$ Hz frequency 
range for two energy bands ($2 - 13$ and $13 - 60$ keV) using short 
stretches of data. For the observations with a high variability 
separate power density spectra were produced for the parts of 
observation divided according to the level of the source flux. 
The resulting spectra were logarithmically rebinned when necessary 
to reduce scatter at high frequencies. The power density spectra were 
normalized to squared fractional {\it rms}. The white-noise level 
due to Poissonian statistics corrected for the dead-time effects was 
subtracted.

To study the evolution of the basic timing parameters of GRS 1915+105
within the individual observations with a relatively high level of 
variability we generated power density spectra of the source 
accumulated in the $16 - 80$ {\it s} time intervals according to the 
procedure described above.

Typical broad-band power density spectra of GRS 1915+105 during 
high luminosity 'flaring' state (HLS) prior to Oct. 23, 1996, 
transition from the HLS to low luminosity state (LLS) and LLS are 
shown in Figure \ref{power_general}. For the 'flaring' state two 
spectra corresponding to the highest ('flare') and lowest ('quescence') 
levels of source luminosity are displayed (Figure \ref{power_general},
{\it top left panel}).

The character of PDS evolution during the source state transitions 
between October 23 and November 28 is very complicated. Fast changes 
in the form of the power spectrum, correlated with spectral and 
luminosity changes on the various time scales (from $\sim$ several 
hours to $\sim$ week) were detected (Figure \ref{power_general}, 
{\em right upper, left lower panel}). Inspite of 
drastic differences in the form of the broad-band continuum,
relatively narrow QPO peak at frequencies $\sim 2 - 10$ Hz has 
been found to be common for all observations covering this period. 
Basing on the results of the time-resolved PDS analysis of the
individual observations with a high level of flux variability, the 
correlation between the value of the QPO centroid frequency and 
total X-ray flux and a bolometric flux of the soft spectral 
component was established (Figure \ref{flux_f}).

Beginning on November 28, when the source reached the low luminosity 
state, the form of the power density spectrum is nearly flat between 
0.1 and 1.0 Hz, changing its slope to a value of $\sim - (1.0 - 2.5)$ 
in the $1 - 15$ Hz and $\sim 2.0 - 3.0$ in the $15 - 50$ Hz range. 
For some observations the power density spectrum shows a notable rise 
(sometimes the extra component is flat-toped or has a power law form) 
towards lower frequencies between 0.01 and 1.0 Hz. The most outstanding 
features in the GRS 1915+105 power density spectrum are the strong, 
relatively narrow ($\Delta f/ f \sim 0.1 - 0.3$) QPOs placed near 
the breakpoint in the slope of the broad-band continuum. These QPOs 
were noted for showing harmonics with a strength that decreases with 
increasing of the QPO centroid frequency. Positive correlation between 
the QPO centroid frequency and total X-ray luminosity of the source 
on the $\sim$ several hours time scale was detected again 
(Figure \ref{flux_f}), however this type of correlation breaks on 
the longer time scales.

We fitted the power density spectra in the $0.05 - 50$ Hz frequency range 
to analytic model using $\chi^{2}$ minimization technique to trace the 
evolution of their main parameters with time during transitions and low 
luminosity state. The data of Oct., 10, 13, 15, 25 and Nov., 19 were 
excluded from the analysis because of the complicated shape of the PDS 
caused by the existence of several distinct types of the PDS due to 
extremely high level of source variability during these observations.  

\subsection*{Analytic approximation}
To approximate the broad-band PDS continuum and quantify its
characteristics, we used the sum of up to three band-limited 
noise components, expressed with flat-toped broken power law 
functions, i. e.:

$$P({\it f})=\cases{{\it A},&$f<f^{br}$;\cr 
              {\it A(f/f^{br})^{-\alpha}},&$f>f^{br}$,\cr}$$ 
where $f_{br}$ is a BLN characteristic break frequency; and 
in some cases add a power law (PL) component:

$P({\it f})= B f^{- \alpha}$,\\ where {\it A} 
and $B$ are in units of {\it (rms/mean)$^{2}$}. Up to four Lorentzian 
peaks were used to model a QPO features. In Figure \ref{model} the 
schematic presentation of the model is shown. Simple fits with
the Lorentzian shapes have shown that regardless of the continuum model 
type centroid frequencies and widths of the QPO peaks are harmonically 
related. Taking this fact into account in order to reduce the total number 
of model parameters we fixed centroid frequencies and widths of QPO 
components to harmonic ratios. 

The best-fit values of the model parameters are summarized in Tables 
\ref{time_fit}, \ref{time_fit_1} and \ref{time_con}, \ref{time_con_1}. 
Parameter errors and upper limits correspond to $1 \sigma$ and $2 \sigma$ 
confidence levels respectively. This model approximates the 
data reasonably well, as indicated by the values of reduced $\chi^{2}$ 
of the fit.

\subsection*{Correlations between timing parameters}
In order to summarize the results of analysis of the source power
density spectra during the low luminosity state and state 
transitions, the main best-fit parameters of the PDS are shown as
a functions of the fundamental QPO peak frequency for the soft ($2 -
13$ keV) and hard ($13 - 60$ keV) energy bands (Figure \ref
{tot_time_soft}, \ref{tot_time_hard}). ({\it Open circles} 
correspond to the observations covering the transition from the high 
luminosity state (HLS) to the low luminosity state (LLS) prior to 
Nov. 28, 1996 (MJD 50415); {\it solid circles} correspond to the 
period of LLS (Dec. 1996 -- Apr. 1997)).

{\it Soft energy band (2 - 13 keV)} As it is clearly seen from Figure 
\ref{tot_time_soft}, the QPO centroid frequency is strongly 
anticorrelated with the value of the total fractional {\it rms} 
(except for the cases when the contribution of the power law (PL) 
noise component to the total variability is noticeable, which is 
probably due to presence of the prominent soft component in the 
energy spectrum (Nov., 7 and Apr., 2 observations)) and with the 
{\it rms} of the band-limited noise. The same tendency is hold for 
the value of fractional {\it rms} of the fundamental QPO peak: it 
decreases with the increase of the QPO frequency. In addition, there 
is a close relation between the positions of the QPO peaks and
characteristic breaks in the BLN continuum 
(Figure \ref{tot_time_soft}, {\it top panels}). 

{\it Hard energy band (13 - 60 keV)} Most of the correlations, noted
for the soft energy band are also observed for the data in the hard 
band, except for the total {\it rms} and BLN {\it rms}, which do not 
show any correlation with the QPO centroid frequency (Figure 
\ref{tot_time_hard}). 

\subsection*{Correlations between spectral and timing parameters} 
We have analyzed the relation between the spectral and temporal
properties of GRS 1915+105. During the low luminosity state and state 
transitions the change of the QPO centroid frequency is correlated 
with the change of the spectral parameters, derived using our 
simplified spectral model. In figure \ref{rms_flux} the relationship 
between the total fractional {\it rms} integrated over $0.05 - 50$ Hz 
frequency range and X-ray flux ($3 - 20$ keV) for the soft ($2 -
13$ keV) and hard ($13 - 60$ keV) energy bands is presented. 
As the contribution of the soft spectral component to the $13 - 60$
keV flux is small, we can examine the properties of the hard spectral 
component alone using variability analysis of the data in this range. 
As it is seen from figure \ref{rms_flux}, the level of the 
band-limited noise continuum {\it rms} in the $13 - 60$ keV energy 
domain is strongly correlated with the total X-ray flux and, in 
particular, with the hard spectral component flux (Figure 
\ref{rms_flux_hard}), which makes a main contribution to the total
X-ray flux. {\it There is a clear trend of rising the QPO frequency 
with the rise of the soft component flux (Figure \ref{soft_freq}). 
It should be noted that this type of correlation holds on the wide 
range of time scales (from $\sim$ seconds to $\sim$ months).} Further 
discussion of the frequency/spectrum correlations are presented 
in Trudolyubov \etal 1998.

\section*{Summary}
We have presented the results of the RXTE observations of the 
GRS 1915+105 during its low luminosity state in November, 1996 
-- April, 1997. 

\begin{figure}
\vspace{-3.0cm}
\epsfxsize=13.0cm
\epsffile{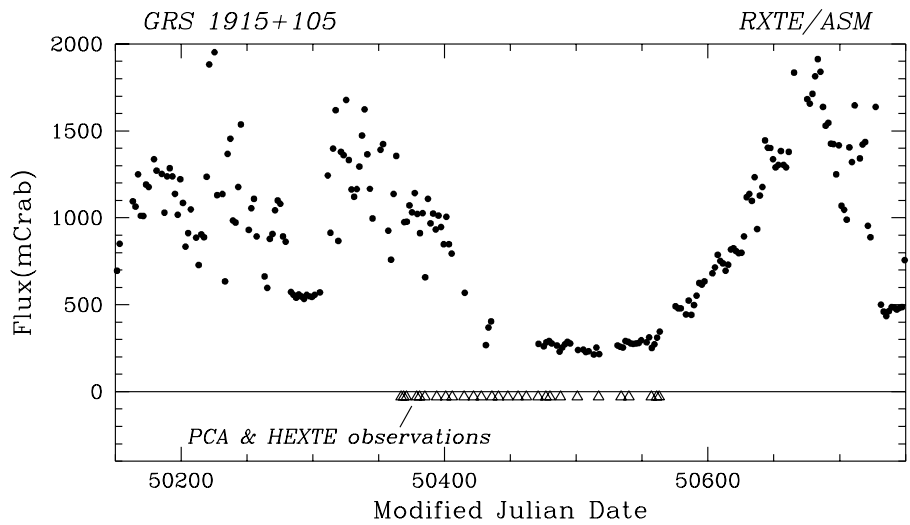}
\caption{\small The {\it RXTE}/ASM light curve of GRS 1915+105 
in the $2 - 12$ keV energy band. The dates of the pointed PCA 
and HEXTE observations are marked with triangles. These particular 
observations were chosen in order to cover the period of extended 
low luminosity state (hereafter LLS) of the source and the transition 
from and to higher luminosity states (hereafter HLS). 
\label{lightcurve}}
\end{figure}

\begin{figure}
\vspace{-3.0cm}
\epsfxsize=13.0cm
\epsffile{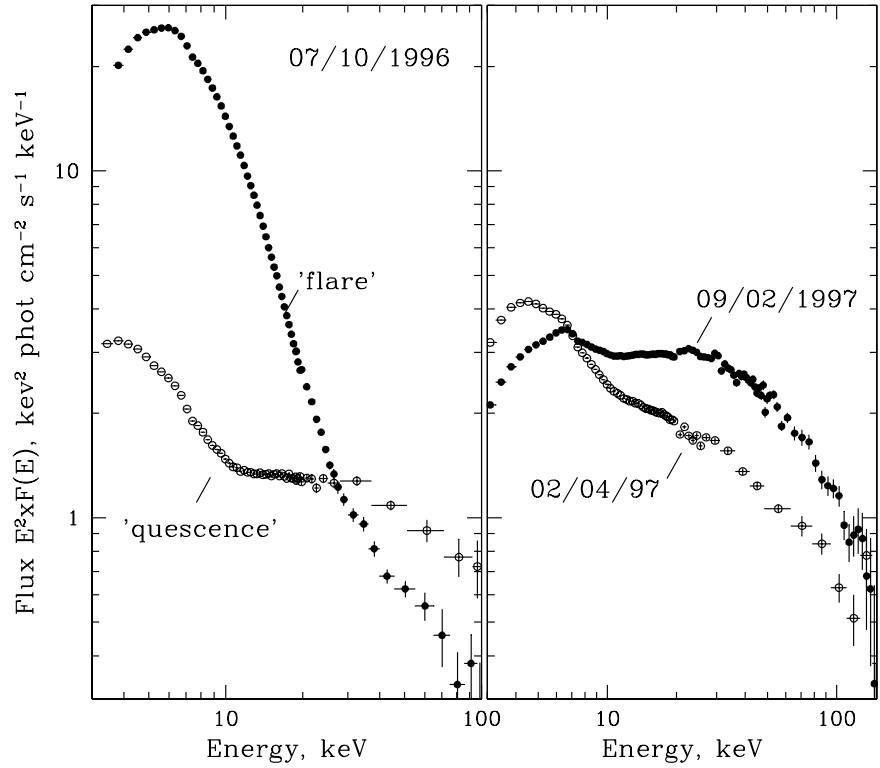}
\caption{\small Typical broad-band energy spectra of GRS 1915+105 in the 
high luminosity 'flaring' ({\it left panel}) and low luminosity
({\it right panel}) states (PCA and HEXTE data). Open and solid circles 
in the {\it left panel} represent the data for the high (`outburst') and 
low (`quescence') levels of source luminosity respectively. 
\label{spectra}}
\end{figure}

\begin{figure}
\vspace{-3.0cm}
\epsfxsize=13.0cm
\epsffile{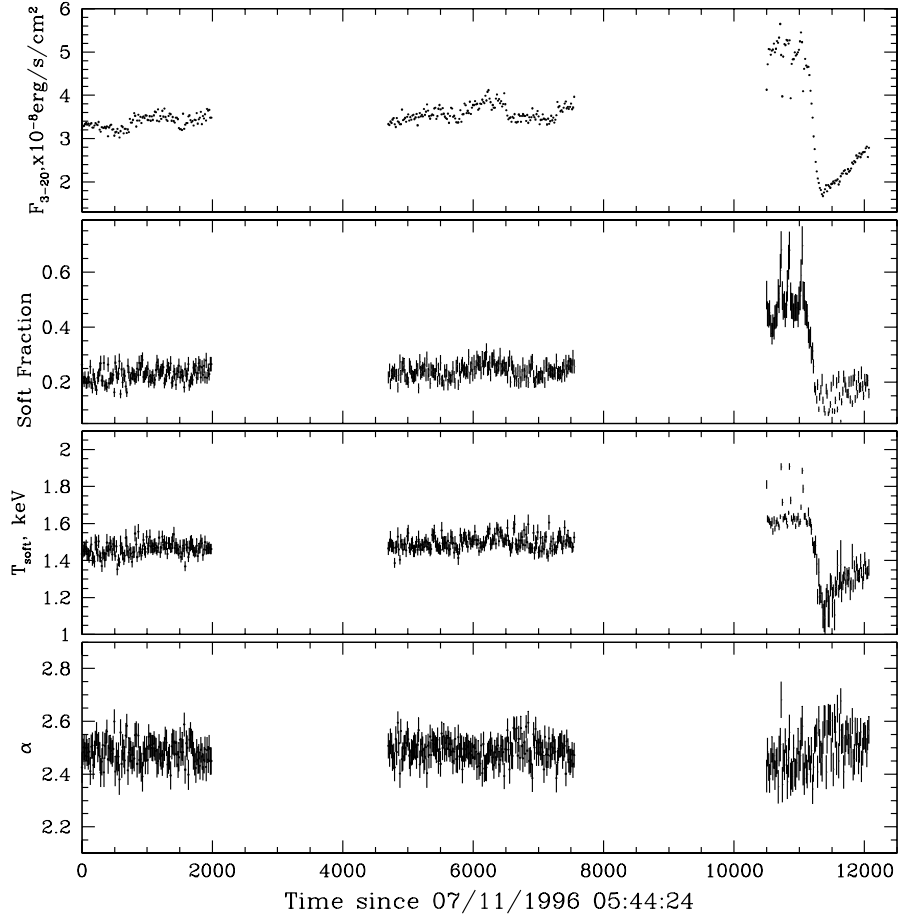}
\caption{\small The evolution of the GRS 1915+105 spectral parameters for 
the November 7, 1996 observation (multicolor disk black body 
model plus power law approximation, the equivalent hydrogen column 
density was fixed at the value of 5$\times 10^{22}$ {\it $cm^{-2}$})
($3 - 20$ keV energy range, PCA data). 'Soft fraction' denotes 
the ratio of the soft component luminosity to the total 
luminosity in the $3 - 20$ keV energy range. Each point corresponds to
the 16 {\it s} integration time. \label{pca_071196}}
\end{figure}

\begin{figure}
\vspace{-3.0cm}
\epsfxsize=13cm
\epsffile{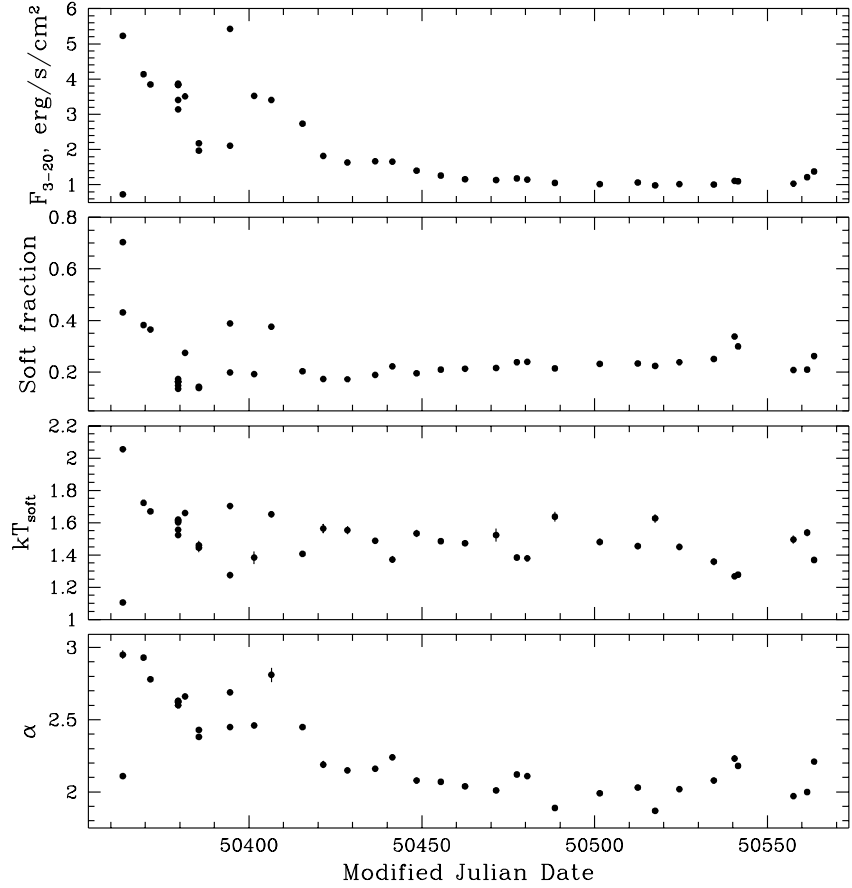}
\caption{\small Evolution of the GRS 1915+105 spectral parameters (multicolor
disk black body model plus power law approximation)($3 - 20$ keV
energy range, PCA data). 'Soft fraction' denotes the ratio of the soft
component luminosity to the total luminosity in the $3 - 20$ keV
energy range. ({\it Note: for MJD 50363, 50369, 50371, 50381, 50406 
observations characterized by a high level of variability the results
of fitting presented by several points corresponding to the 
different levels of source luminosity}).
\label{pca_general}}
\end{figure}

\begin{figure}
\vspace{-3.0cm}
\epsfxsize=13cm
\epsffile{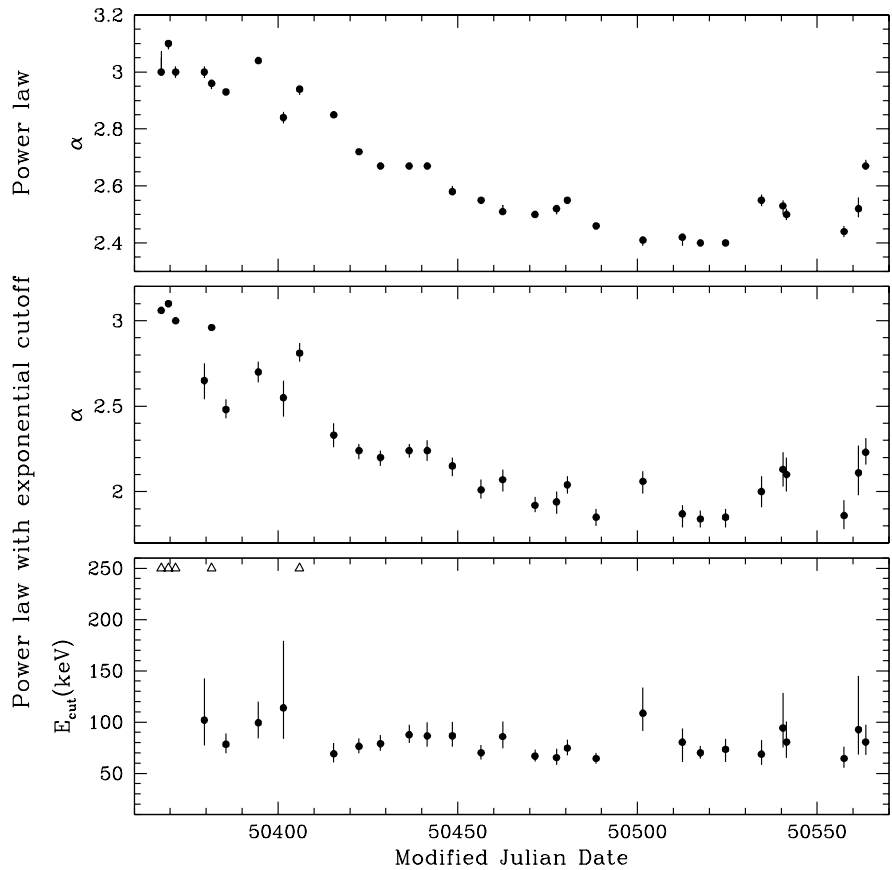}
\caption{\small The evolution of the hard X-ray spectral parameters ({\em upper
panel} -- power law model approximation; {\em middle and bottom panel} --
approximation by power law with exponential cutoff) of GRS 1915+105 during 
its low luminosity state and state transitions in 1996/1997 ($20 - 150$ keV
energy band, HEXTE data). Note: open triangles ({\it bottom panel}) 
show HEXTE upper energy boundary (250 keV) in the cases when the high 
energy cutoff is not detectable within HEXTE energy range
\label{hexte_general}} 
\end{figure}

\begin{figure}
\vspace{-3.0cm}
\epsfxsize=13.0cm
\epsffile{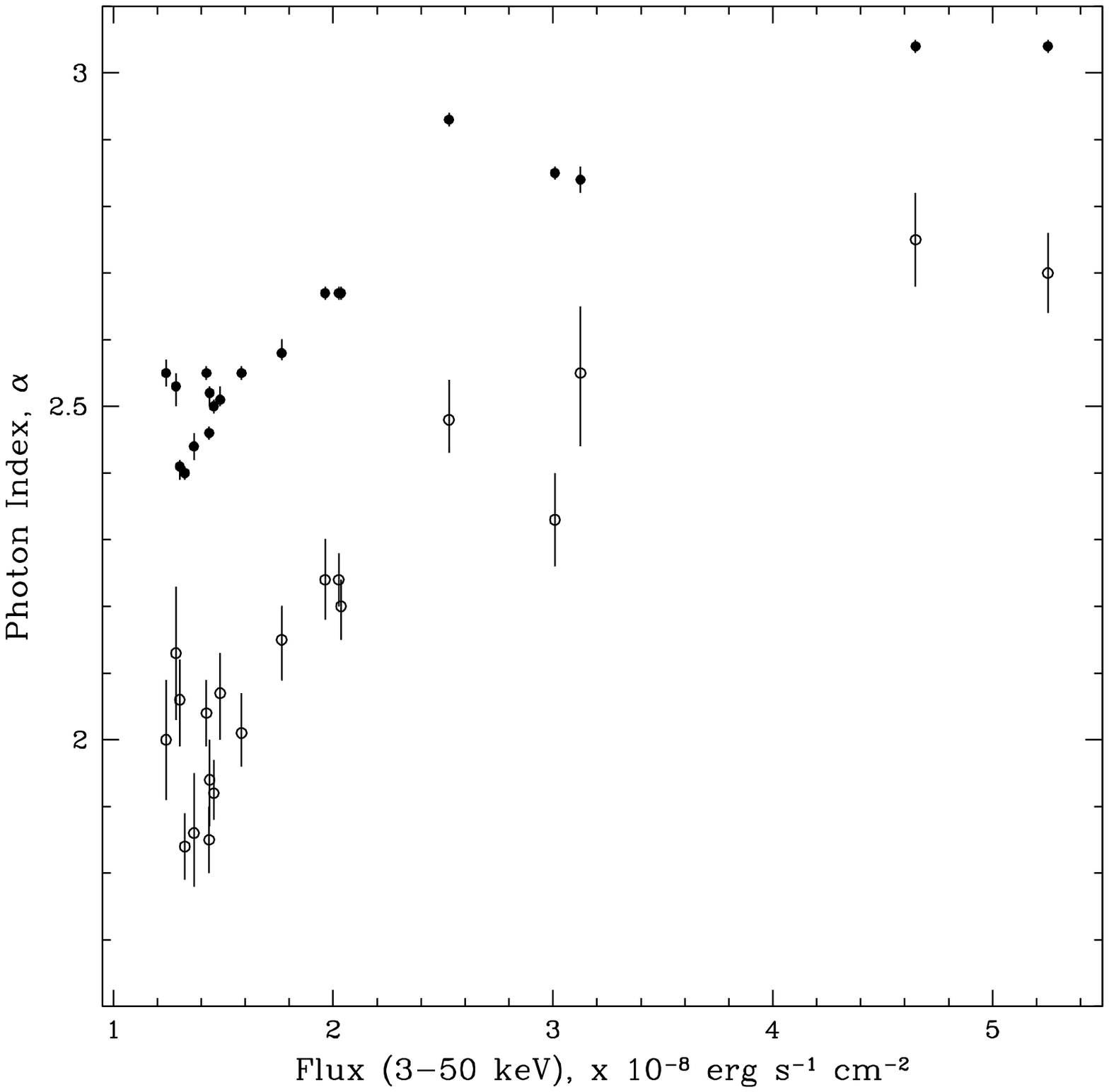}
\caption{\small The slope of the high energy part of GRS 1915+105 spectrum 
(simple power law approximation -- {\it solid circles}; approximation
by power law with exponential cutoff -- {\it hollow circles}) vs. 
$3 - 50$ keV luminosity of the source (PCA and HEXTE data) 
\label{alpha_flux}}
\end{figure}

\begin{figure}
\vspace{-3.0cm}
\epsfxsize=14.0cm
\epsffile{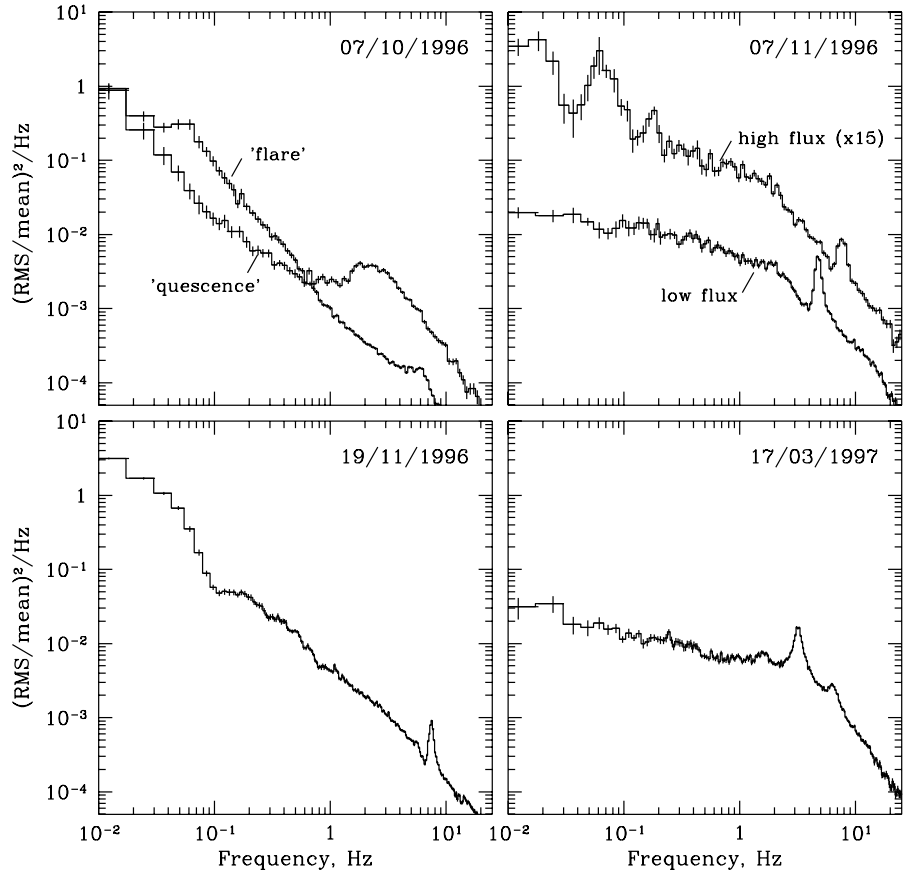}
\caption{\small Typical power density spectra of GRS 1915+105 during low
luminosity state and state transitions: {\it left upper panel} -- 
during {\it high} 'flaring' state; {\it right upper panel} -- 
transition (when the contribution of the soft spectral
component was relatively small); {\it left lower panel} -- transition 
(when the contribution of the soft spectral component is sufficiently 
higher); {\it right lower panel} -- nominal low luminosity state. 
\label{power_general}}
\end{figure}

\begin{figure}
\vspace{-3.0cm}
\epsfxsize=14.0cm
\epsffile{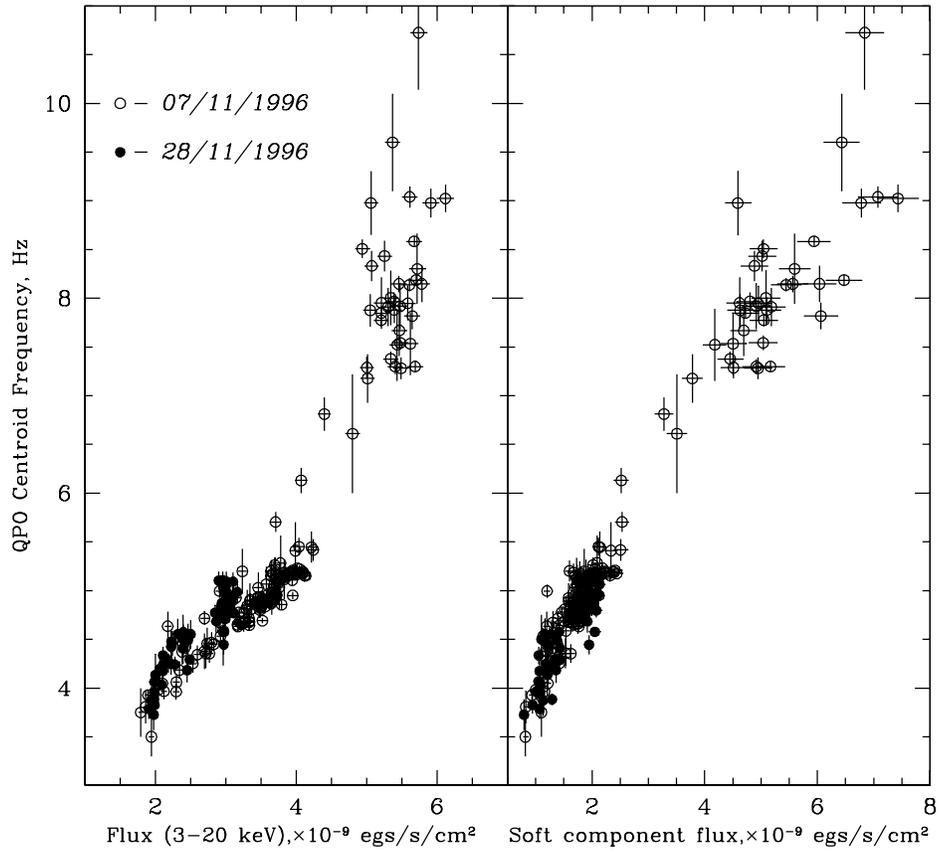}
\caption{\small The evolution of the QPO centroid frequency in 
GRS 1915+105 power density spectrum on the source $3 - 20$ keV 
flux (corrected for the interstellar absorption) and a bolometric 
flux of the soft spectral component during November 7 ({\it open
circles}) and November 28, 1996 ({\it solid circles}) {\it RXTE} 
pointed observations. (Each point represents the data averaged over 
$16 - 48$ {\it s} time intervals; PCA data). \label{flux_f}}
\end{figure}

\begin{figure}
\vspace{-3.0cm}
\epsfxsize=13.0cm
\epsffile{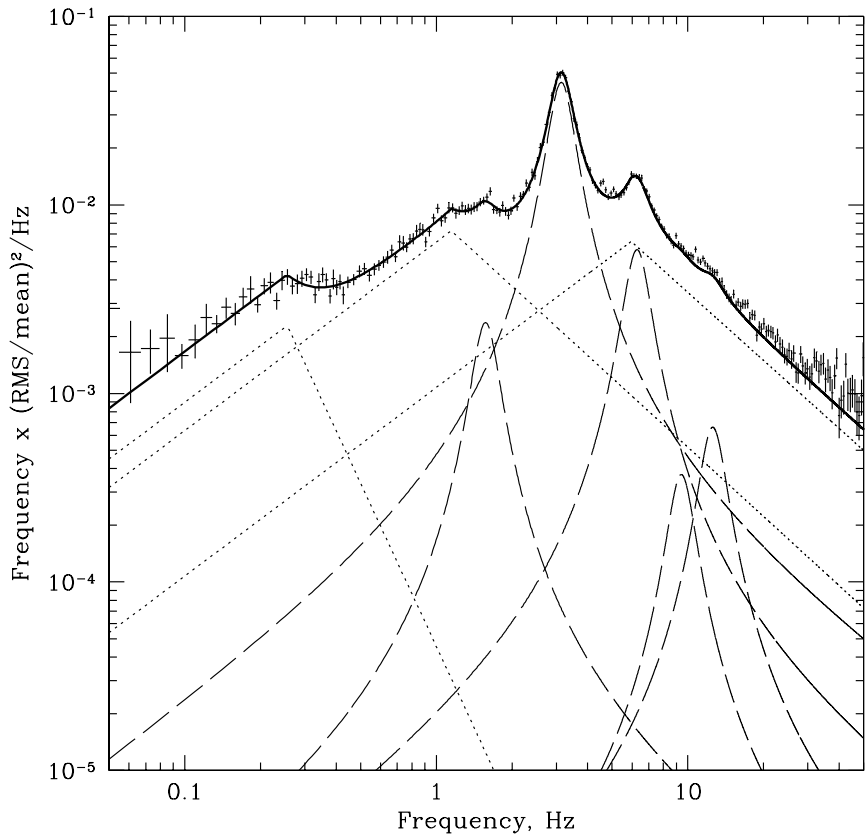}
\caption{\small Schematic presentation of the model used for the approximation 
of the broad-band power density spectra during the low luminosity state and
state transitions ({\em thick solid line}). The contributions of band 
limited noise (BLN) components, Lorentzian components are shown by 
{\em dotted lines} and {\em long-dashed lines} respectively. The data 
for the Dec. 19, 1996 observation are shown for the comparison 
(PCA data, $2 - 13$ keV energy range). \label{model}}
\end{figure}

\begin{figure}
\vspace{-3.0cm}
\epsfxsize=13.0cm
\epsffile{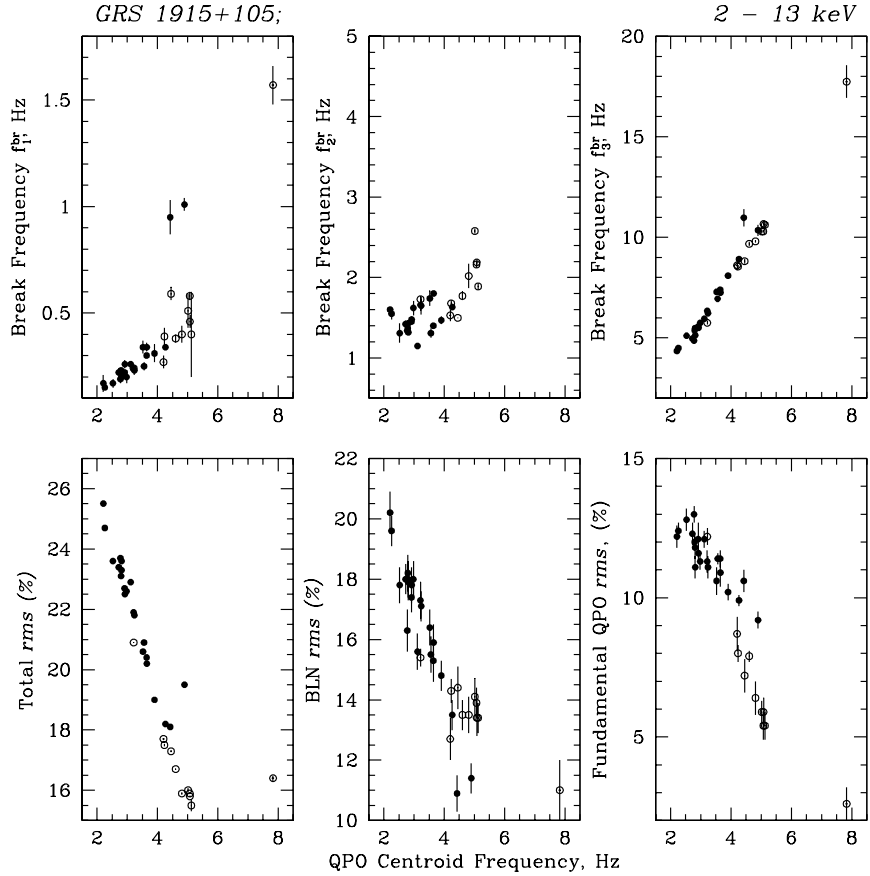}
\caption{\small Main best-fit parameters of the source power density 
spectrum (see Tables \ref{time_fit}, \ref{time_fit_1}, 
\ref{time_con}, \ref{time_con_1}) as a functions of the fundamental 
QPO peak frequency ($2 - 13$ keV energy range). {\it Open circles} 
correspond to the observations covering the transition from the high 
luminosity state (HLS) to the low luminosity state (LLS) prior to 
Nov. 28, 1996 (MJD 50415); {\it solid circles} correspond to the 
period of LLS (Dec. 1996 -- Apr. 1997). 
\label{tot_time_soft}} 
\end{figure}

\begin{figure}
\vspace{-3.0cm}
\epsfxsize=13.0cm
\epsffile{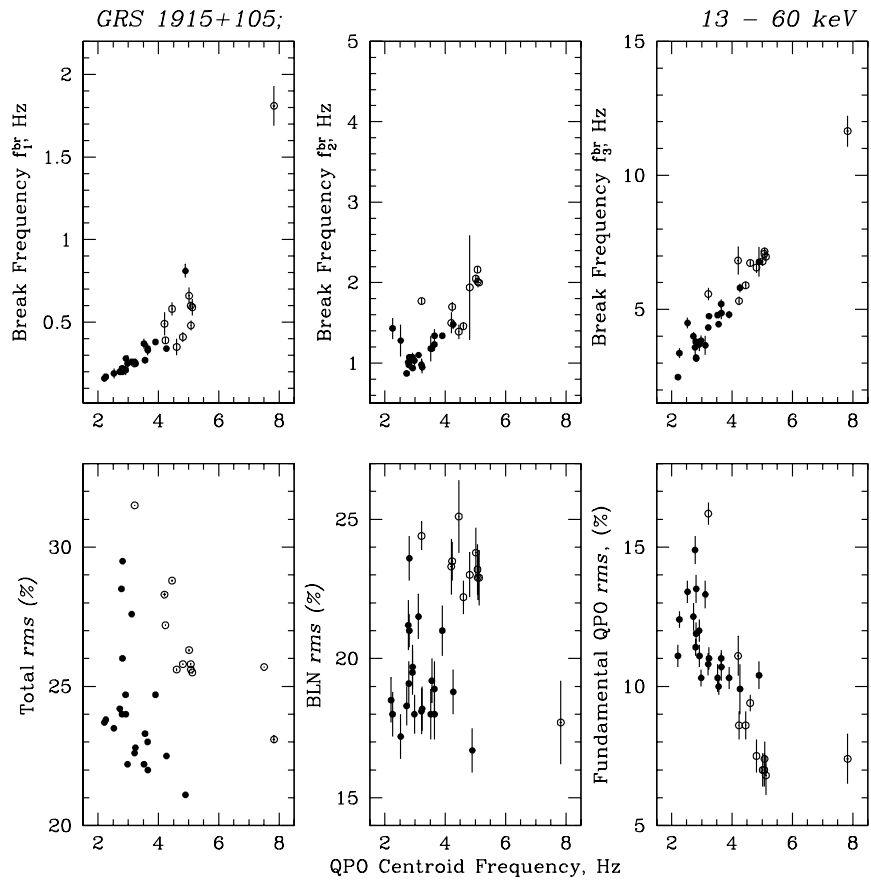}
\caption{\small The same as Figure \ref {tot_time_soft} but for the 
$13 - 60$ keV energy band. \label{tot_time_hard}} 
\end{figure}

\begin{figure}
\vspace{-3.0cm}
\epsfxsize=13.0cm
\epsffile{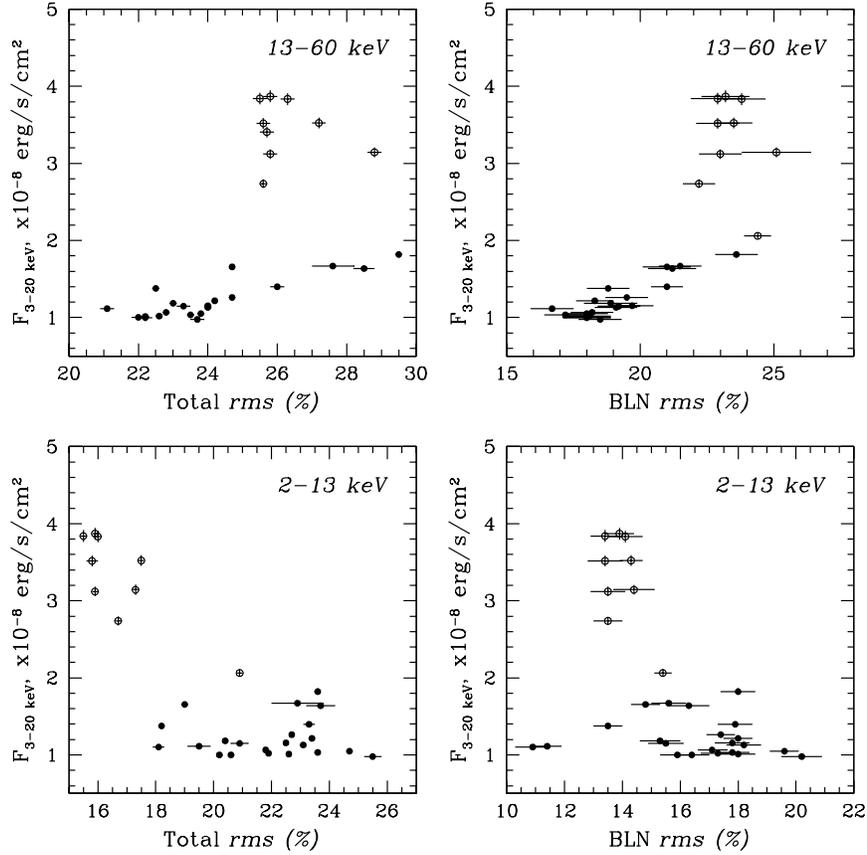}
\caption{\small Total fractional {\it rms} and band-limited component 
{\it rms} in the 2 -- 13 keV and 13 -- 60 keV energy intervals 
integrated  $0.05 - 50$ Hz frequency range vs. total 3 -- 20 keV 
X-ray flux. {\it Open circles} correspond to the observations 
covering the transition from the high luminosity state (HLS) to the 
low luminosity state (LLS) prior to Nov. 28, 1996 (MJD 50415); 
{\it solid circles} correspond to the period of LLS (Dec. 1996 -- 
Apr. 1997). \label{rms_flux}} 
\end{figure}

\begin{figure}
\vspace{-3.0cm}
\epsfxsize=13.0cm
\epsffile{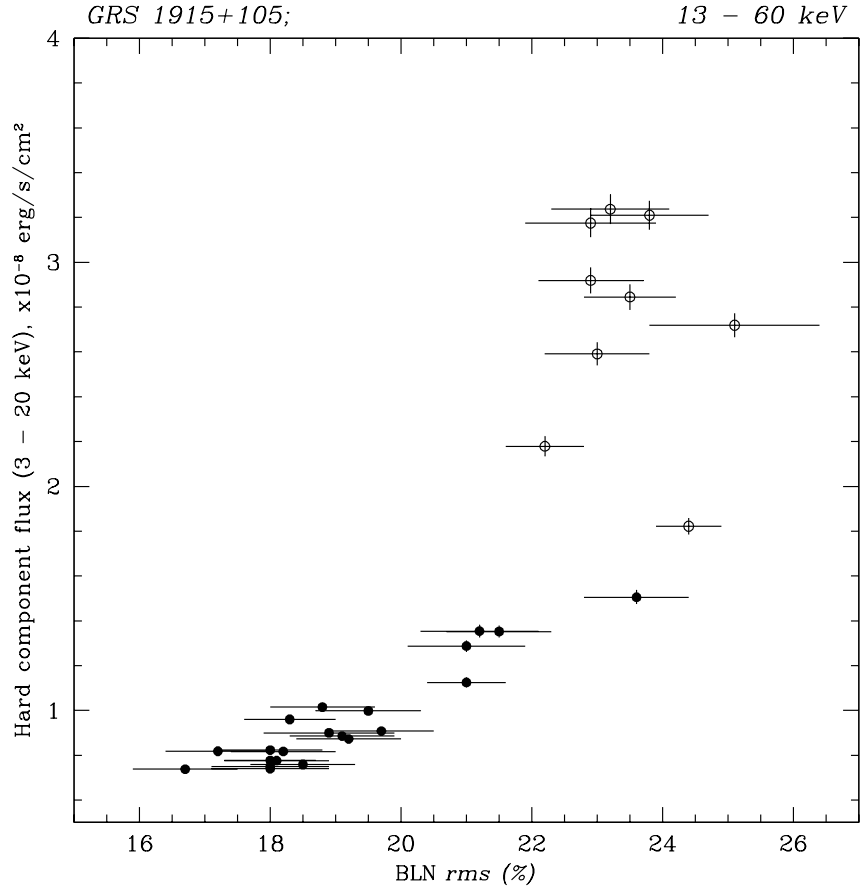}
\caption{\small Total fractional {\it rms} and band-limited component 
{\it rms} in the 13 -- 60 keV energy interval integrated 0.05 -- 50 
Hz frequency range vs. 3 -- 20 keV hard spectral component flux. 
{\it Open circles} correspond to the observations covering the 
transition from the high luminosity state (HLS) to the low luminosity 
state (LLS) prior to Nov. 28, 1996 (MJD 50415); {\it solid circles} 
correspond to the period of LLS (Dec. 1996 -- Apr. 1997). 
\label{rms_flux_hard}} 
\end{figure}

\begin{figure}
\vspace{-3.0cm}
\epsfxsize=15.0cm
\epsffile{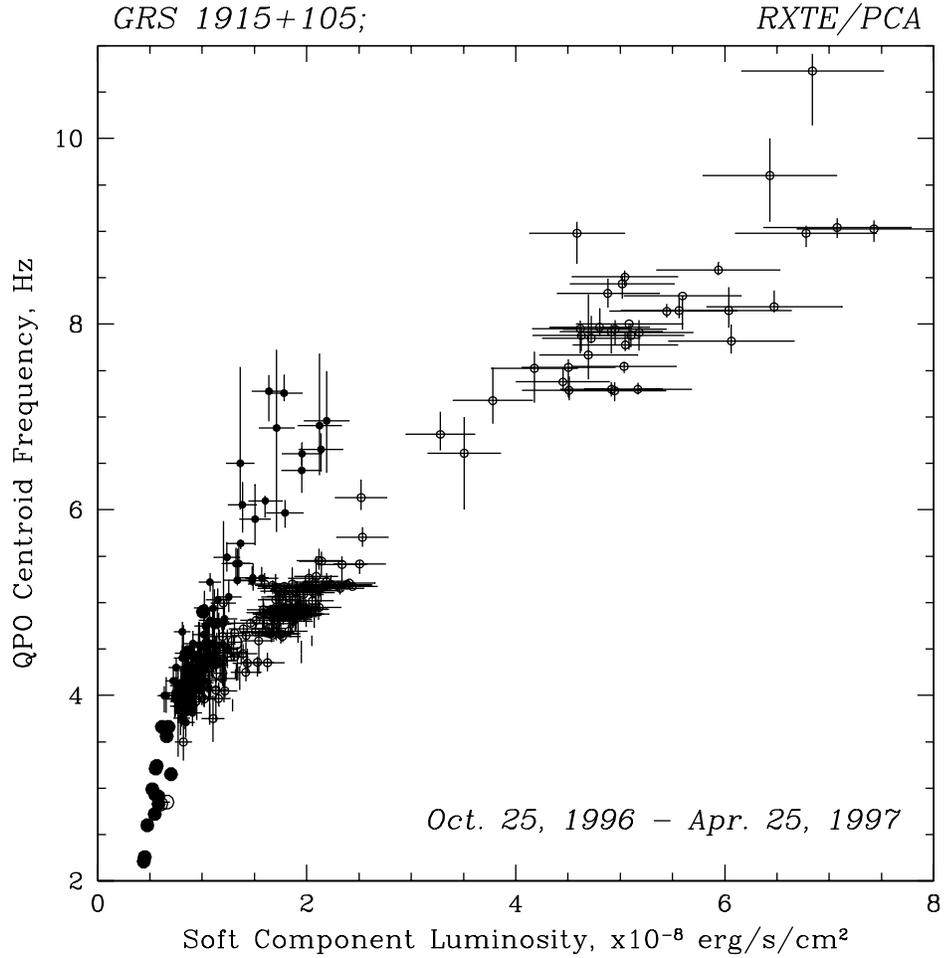}
\caption{\small The dependence of the QPO centroid frequency on the 
estimated bolometric luminosity of the soft component obtained 
from the spectral fitting with our simplified model (multicolor 
disk black body plus power law) for the November, 1996 -- April, 
1997 low luminosity state and state transitions of GRS 1915+105. 
{\it Open circles} correspond to the observations, covering 
the period of transition from HLS to LLS prior to November 28, 1997 
(MJD 50415); {\it filled circles} correspond to the period of low luminosity
state and following rise of the source flux. {\it Large circles} 
represent the averaging over the whole observation; the results for 
the observations with high level of variability averaged over 
$32 - 80$ {\it s} time intervals are presented by {\it small circles}. 
\label{soft_freq}} 
\end{figure}

This period is characterized by the drop of the source flux in 
the standard X-ray band by a factor of $\sim 2-5$ with respect 
to the bright flaring state (Figure \ref{lightcurve}). Contrary 
to the high luminosity state (HLS) with its high level of 
variability, the evolution 
of the source flux during low luminosity state can be described 
as a relatively smooth decline to the lowest level during first 
$\sim 100$ days followed by similar-fashioned slow rise on the 
same time scale.  

The broad-band X-ray spectrum of GRS 1915+105 can be generally 
described by composition of the soft component and an extended 
high energy tail (Figure \ref{spectra}, {\it right panel}). 
In order to characterize the general character of the source 
spectral evolution we used a composition of the multicolor 
disk blackbody model and a power law model with exponential 
cut-off. The contribution of the soft component, dominating the 
source luminosity in the HLS prior to transition, dropped to 
$< 25 \%$ in the $3 - 20$ keV energy band. In general 
the evolution of GRS 1915+105 broad-band spectrum during low
luminosity state and state transitions can be characterized by the 
decrease of the soft component flux accompanied by 
gradual flattening of the hard spectral component with decrease 
of the total source intensity, followed by rise of the soft 
component flux and hard component steepening as the overall source 
luminosity become increasing again. 

The temporal properties of GRS 1915+105 during LLS were also 
quite different from those in the HLS. The
overall shape of the source power density spectrum in the $0.01 - 50$ 
Hz frequency is roughly presented by the composition of the flat-toped
band-limited (BLN) and a power law (PL) noise components. Strong, 
relatively narrow ($\delta f / f \sim 0.2$) QPOs with a centroid 
frequency at about $2 - 10$ Hz located near the break of the BLN 
continuum slope is found to be a generic feature of the 
GRS 1915+105 power density spectrum. In addition, the QPO peaks 
are noted for showing harmonic content. Close relation between the 
main parameters describing the power density spectrum of the source 
(QPO centroid frequency, characteristic BLN break frequency, 
broad-band {\it rms}) may hint on the general character of the 
band-limited noise and QPO production.

{\it The most striking is the established close relation between the 
evolution of the spectral and timing parameters of the source. 
The change of the QPO centroid frequency is correlated with the 
change of the spectral parameters, 
derived using our simplified spectral model. In particular, there 
is a clear trend of rising the QPO frequency with the rise of the 
soft component flux (Figure \ref{soft_freq}). It should be noted 
that this type of correlation holds on the wide range of time scales 
(from $\sim$ seconds to $\sim$ months).}   

The complex of GRS 1915+105 X-ray properties is very similar to 
that of some Galactic black hole candidates observed in the so-called 
{\it 'intermediate'} state during their {\it high-to-low} and {\it 
low-to-high} state transitions (Cygnus X-1 (Belloni \etal 1996, 
Cui \etal 1997); GX 339-4 (Mendez $\&$ Van der Klis 1996); GS 1124-68 
(Miyamoto \etal 1994, Takizawa \etal 1996); GRO J1655-44 
(Mendez \etal 1998)). In fact, there are 
several general properties of GRS 1915+105 showing that this state 
of the source does not match neither 'canonical' {\it low} nor {\it high} 
state and probably corresponds to the transition between these states:\\ 
1) the overall X-ray luminosity in the low luminosity state 
$L_{LLS} \sim 2 \times 10^{38} erg/s$ is sufficiently lower than 
in the high luminosity state $L_{HLS} > 10^{39} erg/s$, which was 
observed earlier and had the properties typical for the {\it high} 
state;\\
2) contrary to the typical {\it low} state energy spectrum of the 
source shows clear evidence for the soft component, but its contribution 
to the total X-ray luminosity is small ($< 30 \%$) instead of $\sim 
70 - 90 \%$ in the case of {\it high} state;\\
3) the power density spectrum also differs from the typical power density 
spectra in the {\it low} and {\it high} states: contrary to the
'canonical' {\it soft} state it shows the presence of the prominent 
band-limited noise component, while the frequency of the characteristic
peak of the BLN continuum ($\sim 2 - 10$ Hz) is one order of magnitude
higher than usually expected for the typical low state of Galactic
black hole candidates.  

Basing on the results of {\it RXTE} observations we can conclude that 
during this low luminosity episode inspite of some peculiarities, the
major X-ray properties of GRS 1915+105 were similar to that of
Galactic black hole candidates in the {\it transition} or
'intermediate' state. The observed trends of hardening the energy
spectra, rise of the BLN {\it rms}, decrease of the QPO frequency with
the decrease of the source luminosity seem to suggest that further
decrease of the luminosity (below the values observed during November,
1996 -- April, 1997) would put the source to the canonical low state
established for the Galactic black hole candidates. It is likely that
the source was observed in the 'intermediate' state simply because the
luminosity (mass accretion rate) did not drop sufficiently enough for
the source to reach canonical {\it low/hard} state.

This research has made use of data obtained through the High 
Energy Astrophysics Science Archive Research Center Online Service, 
provided by the NASA/Goddard Space Flight Center.

This work was supported in part by the grants RBRF grant 96-02-18588 and
INTAS grant 93-3364-ext. S. Trudolyubov was partially supported by 
grant of the International Science Foundation. 

\subsection*{References}
{\small
Belloni, T., Mendez, M., Van der Klis, M., Hasinger, G., Lewin,
W. H. G., $\&$ Van Paradijs, J. 1996, ApJ, 472, L107 \\
Belloni, T., van der Klis, M., Lewin, W. H. G., Van Paradijs,
J., Dotani, T., Mitsuda, K., Miyamoto, S. 1997, A$\&$A, 322, 857 \\
Belloni, T., Mendez, M., King, A. R., Van der Klis, M., $\&$ Van 
Paradijs, J. 1997{\em a}, ApJ, 479, L145 \\
Belloni, T., Mendez, M., King, A. R., Van der Klis, M., $\&$ Van 
Paradijs, J. 1997{\em b}, ApJ, 488, L109 \\
Bradt, H., Swank, J., $\&$ Rothschild, R. 1993, A$\&$AS, 97,355\\
Chen, X., Swank, J. H., $\&$ Taam, R. E. 1997, ApJ, 477, L41 \\
Cui, W., Zhang, S. N., Focke, W., $\&$ Swank, J. H. 1997, ApJ, 484, 383 \\
Ebisawa, K., White, N. E., $\&$ Kotani, T. 1995, IAU Circ. 6171 \\
Greiner, J., Morgan, E. H., $\&$ Remillard, R. A. 1998, astro-ph/9806323 \\
Jahoda, K. 1997, \\
http://lheawww.gsfc.nasa.gov/users/keith/pcarmf.html \\
Mendez, M., $\&$ Van der Klis, M. 1996, ApJ, 479, 926 \\
Mendez, M., Belloni, T., $\&$ Van der Klis, M. 1998, ApJ, 489, L187 \\
Mirabel, I., F., $\&$ Rodriguez, L. F. 1994, Nature, 371, 46 \\
Miyamoto, S., Kitamoto, S., Iga, S., Hayashida, K., Terada, K. 
1994, ApJ, 435, 389 \\
Mitsuda, K. \etal 1984, PASJ, 36, 741 \\
Morgan, E. H., Remillard, R. A., $\&$ Greiner, J. 1997, ApJ, 482, 993 \\
Paciesas, W. S., Deal, K. J., Harmon, B. A., Zhang, S. N., 
Wilson, C. A., $\&$ Fishman, G. J. 1996, A$\&$AS, 120, 205 \\
Remillard, R. A., $\&$ Morgan, E. H. 1998, astro-ph/9805237 \\
Sazonov, S., Sunyaev, R., Alexandrovich, N., $\&$ Borozdin, K. 1994, 
IAU Circ. 6080\\
Sazonov, S., Sunyaev, R., $\&$ Lund, N. 1996, in Proc. of 
X-ray Conference, eds. Zimmermann, H.U., Truemper, J., 
$\&$ Yorke, H. (MPE Report 263), 187\\
Shakura, N. I., $\&$ Sunyaev, R. A. 1973, A$\&$A, 24, 337 \\
Shimura, T., $\&$ Takahara, F. 1995, ApJ, 445, 780 \\
Stark 1997, \\ 
http://lheawww.gsfc.nasa.gov/users/stark/pca/pcabackest.html \\
Taam, R. E., Chen, X., Swank, J. H. 1997, ApJ, 485, L83 \\
Takizawa, M., Dotani, T., Mitsuda, K., Matsuba, E., Ogawa, M., 
Aoki, T., Asai, K., Ebisawa, K., Makishima, K., Miyamoto, S., 
Iga, S., Vaughan, B., Rutledge, R., $\&$ Lewin, W. H. G. 1996, ApJ, 
489, 272 \\
Tanaka, Y., $\&$ Lewin, W. H. G. 1995, in X-ray Binaries, 
ed. W. H. G. Lewin, J. Van Paradijs, $\&$ E. P. J. van den Heuvel 
(Cambridge: Cambridge Univ. Press), 126 \\
Trudolyubov, S. \etal (in preparation) \\
Zhang, W., $\&$ Jahoda,
K. 1996, \\ http://lheawww.gsfc.nasa.gov/users/keith/deadtime/deadtime.html\\
Zhang, W., Jahoda, K., Swank, J. H., Morgan, E. H., $\&$ Giles, A. B. 
1995, ApJ, 449, 930 \\
}

\begin{table}
\vspace{-2.0cm}
\caption{\small {\it RXTE} observations of GRS 1915+105 in October, 1996 --
May, 1997. \label{obslog}}
\small
\begin{tabular}{cccccc}
\hline
\hline
\noalign{\smallskip}
 & & & & Exposure$^{a}$, s & \\
\noalign{\smallskip}
\hline
MJD & Date, UT & Time, UT (h:m:s) & PCA & HEXTE--A & HEXTE--B \\
\hline
50363 & 07/10/1996 & $05:44:36 - 11:11:13$ & $10944$& $1252$ & $1100$ \\
50369 & 13/10/1996 & $10:03:17 - 14:06:13$ & $7756$ & $1165$ & $1156$ \\ 
50371 & 15/10/1996 & $15:10:39 - 22:30:13$ & $7856$ & $1274$ & $1253$ \\
50379 & 23/10/1996 & $11:53:56 - 18:25:13$ & $9072$ & $1289$ & $1261$ \\
50381 & 25/10/1996 & $11:52:51 - 17:44:13$ & $7216$ & $1327$ & $1288$ \\
50385 & 29/10/1996 & $11:53:13 - 17:29:14$ & $5116$ & $2700$ & $2717$ \\
50394 & 07/11/1996 & $05:42:38 - 09:07:14$ & $7760$ & $2084$ & $2131$ \\
50401 & 14/11/1996 & $02:17:24 - 03:42:13$ & $2944$ & $888$  & $895$  \\
50406 & 19/11/1996 & $02:52:20 - 07:39:13$ & $6368$ & $3890$ & $1285$ \\
50415 & 28/11/1996 & $03:04:13 - 06:16:13$ & $6969$ & $1275$ & $817$  \\
50421 & 04/12/1996 & $23:25:58 - 03:26:13$ & $1952$ & $1130$ & $1120$ \\
50428 & 11/12/1996 & $18:42:12 - 22:42:14$ & $9232$ & $2339$ & $2340$ \\
50436 & 19/12/1996 & $15:47:20 - 19:47:13$ & $8623$ & $2430$ & $2510$ \\
50441 & 24/12/1996 & $22:04:38 - 01:28:13$ & $5888$ & $1939$ & $1978$ \\
50448 & 31/12/1996 & $06:48:53 - 10:19:13$ & $7284$ & $1353$ & $1349$ \\
50455 & 07/01/1997 & $23:50:28 - 04:08:13$ & $3215$ & $2569$ & $2676$ \\
50462 & 14/01/1997 & $01:29:43 - 03:59:14$ & $6250$ & $1275$ & $1314$ \\
50471 & 23/01/1997 & $01:39:50 - 05:51:14$ & $9016$ & $2287$ & $2342$ \\
50477 & 29/01/1997 & $20:57:32 - 01:38:13$ & $9563$ & $2045$ & $1086$ \\
50480 & 01/02/1997 & $21:08:59 - 01:37:13$ & $9010$ & $2788$ & $2796$ \\
50488 & 09/02/1997 & $18:41:59 - 00:01:13$ & $9847$ & $2483$ & $2515$ \\
50501 & 22/02/1997 & $21:14:51 - 23:36:13$ & $5739$ & $1865$ & $1854$ \\
50512 & 05/03/1997 & $21:20:31 - 22:55:13$ & $4564$ & $1792$ & $1801$ \\
50517 & 10/03/1997 & $01:07:20 - 04:27:13$ & $5648$ & $1890$ & $1891$ \\
50524 & 17/03/1997 & $22:10:05 - 01:55:13$ & $7824$ & $3921$ & $3916$ \\   
50534 & 27/03/1997 & $21:43:35 - 23:14:13$ & $3264$ & $990$  & $974$  \\
50540 & 02/04/1997 & $12:35:28 - 14:14:13$ & $3300$ & $988$  & $996$  \\
50557 & 19/04/1997 & $10:41:03 - 14:50:13$ & $2368$ & $2181$ & $2086$ \\
50561 & 23/04/1997 & $03:08:24 - 05:19:13$ & $3520$ & $1093$ & $1064$ \\
50563 & 25/04/1997 & $14:03:56 - 20:56:14$ & $13105$& $3467$ & $2481$ \\
\hline
\end{tabular}
\begin{list}{}{}
\item[$^a$] -- dead time corrected value of used exposure time 
\end{list}
 
\end{table}

\begin{table}
\vspace{-2.0cm}
\caption{\small The evolution of the X-ray flux and the level of 
the short-term variability of GRS 1915+105 during the low luminosity 
state and state transitions in October, 1996 -- April, 1997. For 
observations with a high level of variability the ranges of parameter 
change are displayed. \label{timing}}
\small
\begin{tabular}{ccccc}
\hline
\hline
\noalign{\smallskip}
&  & X-ray Flux$^{*}$ & RMS Intensity & RMS Intensity \\
\noalign{\smallskip}
MJD & Date, UT & $F_{3-20}$, $\times 10^{-8}$ & Variation & Variation \\
\noalign{\smallskip}
&  & (\it erg s$^{-1}$ cm$^{-2}$) & ($\%$)$^{1}$ & ($\%$)$^{2}$ \\
\noalign{\smallskip}
\hline
\noalign{\smallskip}
50379 & 23/10/1996 & $3.538 - 4.650$ & $15.2 - 17.7$ & $26.0 - 27.6$ \\
\noalign{\smallskip}
50385 & 29/10/1996 & $1.967 - 2.182$ & $20.3 - 21.4$ & $31.8\pm2.0$  \\
\noalign{\smallskip}
50394 & 07/11/1996 & $2.111 - 5.427$ & $15.6 - 17.6$ & $24.2 - 28.9$ \\
\noalign{\smallskip}
50401 & 14/11/1996 & $3.523$ & $17.2\pm0.1$ &   $27.2\pm0.2$         \\
\noalign{\smallskip}
50406 & 19/11/1996 & $3.407$ & $33.4\pm0.3$ & $25.7\pm0.2$           \\
\noalign{\smallskip}
50415 & 28/11/1996 & $1.932 - 3.084$ & $16.2 - 17.8$ & $25.1 - 27.1$ \\
\noalign{\smallskip}
50421 & 04/12/1996 & $1.820$ & $23.6\pm0.1$ & $29.8\pm0.1$           \\
\noalign{\smallskip}
50428 & 11/12/1996 & $1.553 - 1.685$ & $21.7 - 24.6$ &  $27.4 - 29.0$\\
\noalign{\smallskip}
50436 & 19/12/1996 & $1.669$ & $22.1 - 24.1$ & $27.7 - 29.1$         \\
\noalign{\smallskip}
50441 & 24/12/1996 & $1.656$ & $19.1\pm0.1$ & $25.4\pm0.1$           \\
\noalign{\smallskip}
50448 & 31/12/1996 & $1.398$ & $23.5\pm0.3$ & $26.7\pm0.2$           \\
\noalign{\smallskip}
50455 & 07/01/1997 & $1.263$ & $22.2 - 23.3$& $25.1 - 25.9$          \\
\noalign{\smallskip}
50462 & 14/01/1997 & $1.155$ & $22.5\pm0.1$ & $24.7\pm0.1$           \\
\noalign{\smallskip}
50471 & 23/01/1997 & $1.131$ & $23.1\pm0.1$ & $24.8\pm0.1$           \\
\noalign{\smallskip}
50477 & 29/01/1997 & $1.183$ & $21.0\pm0.1$ & $24.4\pm0.1$           \\
\noalign{\smallskip}
50480 & 01/02/1997 & $1.148$ & $20.9\pm0.3$ & $24.2\pm0.2$           \\
\noalign{\smallskip}
50488 & 09/02/1997 & $1.049$ & $24.7\pm0.1$ & $24.5\pm0.1$           \\
\noalign{\smallskip}
50501 & 22/02/1997 & $1.012$ & $22.6\pm0.1$ & $23.1\pm0.1$           \\
\noalign{\smallskip}
50512 & 05/03/1997 & $1.067$ & $21.8\pm0.1$ & $23.6\pm0.2$           \\
\noalign{\smallskip}
50517 & 10/03/1997 & $0.978$ & $25.5\pm0.3$ & $24.5\pm0.2$           \\
\noalign{\smallskip}
50524 & 17/03/1997 & $1.020$ & $22.0\pm0.1$ & $23.4\pm0.1$           \\
\noalign{\smallskip}
50534 & 27/03/1997 & $1.002$ & $20.2\pm0.1$ & $22.8\pm0.2$           \\
\noalign{\smallskip}
50540 & 02/04/1997 & $1.114$  & $19.5\pm0.4$& $23.2\pm0.2$           \\
\noalign{\smallskip}
50548 & 10/04/1997 & $1.102$ & $18.1\pm0.2$ & $21.3\pm0.2$           \\
\noalign{\smallskip}
50561 & 23/04/1997 & $1.216$ & $23.4\pm0.1$ & $25.0\pm0.1$           \\
\noalign{\smallskip}
50563 & 25/04/1997 & $1.340 - 1.413$ & $17.0 - 19.0$ & $22.1 - 23.3$ \\
\noalign{\smallskip}
\hline
\end{tabular}

\begin{list}{}{}
\item[$^{*}$] -- value, corrected for interstellar absorption 
\item[$^{1}$] -- $2 - 13$ keV energy band, $0.05 - 50$ Hz frequency range
\item[$^{2}$] -- $13 - 60$ keV energy band, $0.05 - 50$ Hz frequency range
\end{list}

\end{table}

\clearpage

\begin{table}
\vspace{-3.0cm}
\caption{\small The characteristics of the power density spectrum. 
Parameter errors and upper limits correspond to the $1 \sigma$ and $2 
\sigma$ confidence levels respectively. $rms_{BLN}$ represents the 
total {\it rms} of the band-limited components integrated in the 
$0.05 - 50$ Hz frequency range. Observations with sufficiently high 
level of variability are presented by several subsets of data. 
\label{time_fit}}

\tiny
\begin{tabular}{ccccccccc}
\hline
\hline
\noalign{\smallskip}
Energy (keV)&$f_{1}^{br}$(Hz)&$\alpha_{1}$&$f_{2}^{br}$(Hz)&$\alpha_{2}$
&$f_{3}^{br}$(Hz)&$\alpha_{3}$&$rms_{BLN}$($\%$)  \\
\noalign{\smallskip}  
\hline
\noalign{\smallskip}
\multicolumn{8}{c}{\it 23/10/1996}\\
\noalign{\smallskip}
\hline
\noalign{\smallskip}
$2 - 13$ & $0.51^{+0.03}_{-0.08}$ & $2.20^{+0.31}_{-0.34}$ & 
$2.58^{+0.03}_{-0.03}$ & $5.50^{+0.50}_{-1.23}$ & $10.29^{+0.10}_{-0.07}$ & 
$3.10^{+0.09}_{-0.06}$ & $14.1^{+0.6}_{-0.5}$ \\
$13 - 60$ & $0.66^{+0.04}_{-0.05}$ & $2.92^{+0.55}_{-0.60}$ & 
$2.05^{+0.05}_{-0.03}$ & $4.32^{+0.42}_{-0.37}$ & $6.79^{+0.17}_{-0.16}$ & 
$2.27^{+0.09}_{-0.08}$ & $23.8^{+0.9}_{-0.9}$ \\
$2 - 13$ & $0.58^{+0.02}_{-0.02}$ & $2.24^{+0.36}_{-0.13}$ & 
$2.16^{+0.03}_{-0.04}$ & $3.34^{+0.16}_{-0.17}$ & $10.30^{+0.07}_{-0.06}$ & 
$2.88^{+0.06}_{-0.06}$ & $13.9^{+0.5}_{-0.6}$ \\
$13 - 60$ & $0.60^{+0.02}_{-0.02}$ & $2.31^{+0.31}_{-0.32}$ & 
$2.16^{+0.04}_{-0.04}$ & $5.00^{+0.43}_{-1.54}$ & $7.07^{+0.16}_{-0.15}$ & 
$2.42^{+0.10}_{-0.09}$ & $23.2^{+0.9}_{-1.0}$ \\
$2 - 13$ & $0.40^{+0.20}_{-0.04}$ & $2.01^{+0.25}_{-0.17}$ & 
$1.89^{+0.04}_{-0.03}$ & $3.00^{+0.17}_{-0.13}$ & $10.62^{+0.14}_{-0.13}$ & 
$2.97^{+0.09}_{-0.08}$ & $13.4^{+0.5}_{-0.5}$ \\
$13 - 60$ & $0.59^{+0.04}_{-0.05}$ & $2.16^{+0.35}_{-0.13}$ & 
$2.00$ & $4.71^{+0.55}_{-0.40}$ & $6.96^{+0.15}_{-0.21}$ & 
$2.43^{+0.11}_{-0.09}$ & $22.9^{+1.0}_{-1.0}$ \\
$2 - 13$ & $0.59^{+0.03}_{-0.02}$ & $5.03^{+0.21}_{-2.45}$ & 
$1.50$ & $2.37^{+0.10}_{-0.11}$ & $8.81^{+0.13}_{-0.10}$ & 
$3.11^{+0.14}_{-0.09}$ & $14.4^{+0.7}_{-0.6}$ \\
$13 - 60$ & $0.58^{+0.04}_{-0.03}$ & $2.54^{+1.14}_{-0.49}$ & 
$1.39^{+0.09}_{-0.06}$ & $2.00^{+0.63}_{-0.23}$ & $5.90^{+0.14}_{-0.21}$ & 
$2.81^{+0.51}_{-0.27}$ & $25.1^{+1.3}_{-1.5}$ \\
\noalign{\smallskip}
\hline
\noalign{\smallskip}
\multicolumn{8}{c}{\it 29/10/1996}\\
\noalign{\smallskip}
\hline
\noalign{\smallskip}
$2 - 13$ & $0.24^{+0.01}_{-0.01}$ & $1.87^{+0.13}_{-0.10}$ & 
$1.73^{+0.04}_{-0.03}$ & $5.02^{+0.37}_{-1.43}$ & $5.74^{+0.08}_{-0.05}$ & 
$2.57^{+0.04}_{-0.04}$ & $15.4^{+0.3}_{-0.4}$ \\
$13 - 60$ & $0.25^{+0.01}_{-0.01}$ & $1.45^{+0.02}_{-0.01}$ & 
$1.77^{+0.03}_{-0.04}$ & $6.38^{+0.34}_{-1.89}$ & $5.57^{+0.23}_{-0.45}$ & 
$4.08^{+0.32}_{-0.58}$ & $24.4^{+0.5}_{-0.5}$ \\
\noalign{\smallskip}
\hline
\noalign{\smallskip}
\multicolumn{8}{c}{\it 07/11/1996}\\
\noalign{\smallskip}
\hline
\noalign{\smallskip}
$2 - 13$ & $0.40^{+0.04}_{-0.03}$ & $1.97^{+0.22}_{-0.16}$ & 
$2.02^{+0.03}_{-0.15}$ & $3.48^{+0.23}_{-0.30}$ & $9.80^{+0.07}_{-0.12}$ & 
$2.78^{+0.09}_{-0.07}$ & $13.5^{+0.6}_{-0.6}$ \\
$13 - 60$ & $0.41^{+0.02}_{-0.03}$ & $1.63^{+0.19}_{-0.14}$ & 
$1.94^{+0.06}_{-0.07}$ & $4.23^{+0.65}_{-0.54}$ & $6.56^{+0.19}_{-0.23}$ & 
$2.49^{+0.25}_{-0.18}$ & $23.0^{+0.8}_{-1.0}$ \\
$2 - 13$ & $0.46^{+0.02}_{-0.03}$ & $2.05^{+0.16}_{-0.14}$ & 
$2.19^{+0.04}_{-0.04}$ & $3.71^{+0.19}_{-0.18}$ & $10.66^{+0.15}_{-0.10}$ & 
$3.00^{+0.08}_{-0.06}$ & $13.4^{+0.6}_{-0.5}$ \\
$13 - 60$ & $0.48^{+0.03}_{-0.02}$ & $1.83^{+0.17}_{-0.15}$ & 
$2.01^{+0.07}_{-0.04}$ & $4.58^{+0.63}_{-0.37}$ & $7.17^{+0.11}_{-0.18}$ & 
$2.31^{+0.12}_{-0.07}$ & $22.9^{+0.8}_{-0.9}$ \\
$2 - 13$ & $1.57^{+0.05}_{-0.09}$ & $2.20^{+0.04}_{-0.07}$ & 
$-$ & $-$ & $17.74^{+0.81}_{-0.74}$ & $5.68^{+1.13}_{-1.14}$ & 
$11.0^{+1.0}_{-1.3}$ \\
$13 - 60$ & $1.81^{+0.12}_{-0.06}$ & $2.98^{+0.25}_{-0.21}$ & 
$-$ & $-$ & $11.65^{+0.58}_{-0.39}$ & $2.44^{+0.35}_{-0.16}$ & 
$17.7^{+1.5}_{-2.0}$\\
$2 - 13$ & $0.27^{+0.03}_{-0.03}$ & $1.76^{+0.23}_{-0.10}$ & 
$1.53^{+0.05}_{-0.07}$ & $3.05^{+1.19}_{-0.52}$ & $8.60^{+0.21}_{-0.16}$ & 
$2.59^{+0.14}_{-0.12}$ & $12.7^{+0.7}_{-0.8}$ \\
$13 - 60$ & $0.49^{+0.07}_{-0.04}$ & $2.94^{+2.03}_{-1.06}$ & 
$1.50^{+0.13}_{-0.12}$ & $1.99^{+0.51}_{-0.27}$ & $6.83^{+0.51}_{-0.87}$ & 
$2.48^{+0.68}_{-0.32}$ & $23.3^{+1.0}_{-1.4}$ \\
\noalign{\smallskip}
\hline
\noalign{\smallskip}
\multicolumn{8}{c}{\it 14/11/1996}\\
\noalign{\smallskip}
\hline
\noalign{\smallskip}
$2 - 13$ & $0.39^{+0.02}_{-0.04}$ & $1.82^{+0.24}_{-0.15}$ & 
$1.68^{+0.04}_{-0.04}$ & $3.21^{+0.25}_{-0.25}$ & $8.55^{+0.08}_{-0.10}$ & 
$2.79^{+0.12}_{-0.07}$ & $14.3^{+0.4}_{-0.5}$ \\    
$13 - 60$& $0.39^{+0.02}_{-0.02}$ & $1.47^{+0.06}_{-0.04}$ &
$1.70^{+0.06}_{-0.06}$ & $4.82^{+1.04}_{-1.07}$ & $5.32^{+0.16}_{-0.03}$ & 
$3.11^{+0.33}_{-0.30}$ & $23.5^{+0.7}_{-0.6}$ \\
\noalign{\smallskip}
\hline
\noalign{\smallskip}
\multicolumn{8}{c}{\it 28/11/1996}\\
\noalign{\smallskip}
\hline
\noalign{\smallskip}
$2 - 13$ & $0.38^{+0.01}_{-0.02}$ & $1.82^{+0.11}_{-0.09}$ & 
$1.77^{+0.05}_{-0.03}$ & $2.86^{+0.15}_{-0.14}$ & $9.68^{+0.10}_{-0.13}$ & 
$2.79^{+0.08}_{-0.06}$ & $13.5^{+0.5}_{-0.4}$ \\    
$13 - 60$& $0.35^{+0.05}_{-0.05}$ & $1.41^{+0.04}_{-0.03}$ & 
$1.46^{+0.03}_{-0.05}$ & $2.79^{+0.27}_{-0.17}$ & $6.73^{+0.12}_{-0.08}$ & 
$3.39^{+0.40}_{-0.31}$ & $22.2^{+0.6}_{-0.6}$ \\
\noalign{\smallskip}
\hline
\noalign{\smallskip}
\multicolumn{8}{c}{\it 04/12/1996}\\
\noalign{\smallskip}
\hline
\noalign{\smallskip}
$2 - 13$ & $0.20^{+0.01}_{-0.01}$ & $1.91^{+0.26}_{-0.22}$ & 
$1.43^{+0.03}_{-0.02}$ & $1.90^{+0.07}_{-0.05}$ & $5.13^{+0.10}_{-0.10}$ & 
$2.63^{+0.11}_{-0.12}$ & $18.0^{+0.6}_{-0.6}$ \\
$13 - 60$& $0.20^{+0.01}_{-0.01}$ & $1.43^{+0.06}_{-0.03}$ & 
$1.01^{+0.08}_{-0.04}$ & $3.18^{+1.02}_{-0.99}$ & $3.17^{+0.09}_{-0.02}$ & 
$2.59^{+0.15}_{-0.14}$ & $23.6^{+0.8}_{-0.8}$ \\
\noalign{\smallskip}
\hline
\noalign{\smallskip}
\multicolumn{8}{c}{\it 11/12/1996}\\
\noalign{\smallskip}
\hline
\noalign{\smallskip}
$2 - 13$ & $0.19^{+0.02}_{-0.01}$ & $1.77^{+0.25}_{-0.12}$ & 
$1.34^{+0.02}_{-0.03}$ & $2.55^{+0.22}_{-0.37}$ & $4.86^{+0.17}_{-0.10}$ 
& $2.08^{+0.07}_{-0.04}$ & $16.3^{+0.7}_{-0.8}$ \\
$13 - 60$& $0.20^{+0.01}_{-0.01}$ & $1.49^{+0.09}_{-0.08}$ & 
$1.01^{+0.05}_{-0.04}$ & $3.59^{+0.49}_{-0.37}$ & $3.07^{+0.21}_{-0.08}$ 
& $2.01^{+0.14}_{-0.18}$ & $21.2^{+0.9}_{-0.9}$ \\  
\noalign{\smallskip}
\hline
\noalign{\smallskip}
\multicolumn{8}{c}{\it 19/12/1996}\\
\noalign{\smallskip}
\hline
\noalign{\smallskip}
$2 - 13$ & $0.26^{+0.01}_{-0.01}$ & $3.87^{+0.60}_{-0.55}$ & 
$1.15^{+0.03}_{-0.03}$ & $2.22^{+0.21}_{-0.16}$ & $5.96^{+0.08}_{-0.13}$ 
& $2.20^{+0.04}_{-0.04}$ & $15.6^{+0.6}_{-0.7}$ \\
$13 - 60$ & $0.26^{+0.01}_{-0.01}$ & $2.59^{+0.31}_{-0.27}$ & 
$1.10^{+0.02}_{-0.02}$ & $3.84^{+0.68}_{-0.82}$ & $3.66^{+0.35}_{-0.10}$ 
& $2.04^{+0.06}_{-0.06}$ & $21.5^{+0.8}_{-0.9}$ \\     
\noalign{\smallskip}
\hline
\noalign{\smallskip}
\multicolumn{8}{c}{\it 24/12/1996}\\
\noalign{\smallskip}
\hline
\noalign{\smallskip}
$2 - 13$ & $0.31^{+0.04}_{-0.02}$ & $2.71^{+0.81}_{-0.50}$ & 
$1.47^{+0.03}_{-0.05}$ & $2.22^{+0.18}_{-0.19}$ & $8.10^{+0.13}_{-0.11}$ 
& $2.60^{+0.14}_{-0.08}$ & $14.8^{+0.5}_{-0.4}$ \\
$13 - 60$ & $0.38^{+0.01}_{-0.01}$ & $4.15^{+0.73}_{-0.97}$ & 
$1.34^{+0.03}_{-0.02}$ & $3.46^{+0.37}_{-0.35}$ & $4.81^{+0.14}_{-0.05}$ 
& $2.18^{+0.08}_{-0.07}$ & $21.0^{+0.9}_{-0.9}$ \\
\noalign{\smallskip}
\hline
\noalign{\smallskip}
\multicolumn{8}{c}{\it 31/12/1996}\\
\noalign{\smallskip}
\hline
\noalign{\smallskip}
$2 - 13$ & $0.22^{+0.02}_{-0.01}$ & $2.96^{+0.49}_{-0.47}$ & 
$1.32^{+0.02}_{-0.02}$ & $1.73^{+0.05}_{-0.04}$ & $5.48^{+0.06}_{-0.11}$ 
& $2.74^{+0.13}_{-0.15}$ & $17.9^{+0.6}_{-0.5}$ \\
$13 - 60$& $0.22^{+0.01}_{-0.01}$ & $1.80^{+0.24}_{-0.20}$ & 
$1.07^{+0.04}_{-0.04}$ & $5.47^{+1.00}_{-1.98}$ & $3.20^{+0.12}_{-0.06}$ 
& $2.17^{+0.09}_{-0.06}$ & $21.0^{+0.6}_{-0.8}$ \\
\noalign{\smallskip}
\hline
\noalign{\smallskip}
\multicolumn{8}{c}{\it 07/01/1997}\\
\noalign{\smallskip}
\hline
\noalign{\smallskip}
$2 - 13$ & $0.22^{+0.01}_{-0.01}$ & $2.10^{+0.35}_{-0.29}$ & 
$1.45^{+0.03}_{-0.01}$ & $1.73^{+0.05}_{-0.03}$ & $5.57^{+0.05}_{-0.09}$ 
& $2.73^{+0.12}_{-0.12}$ & $17.4^{+0.5}_{-0.5}$ \\
$13 - 60$& $0.21^{+0.03}_{-0.01}$ & $1.46^{+0.18}_{-0.10}$ & 
$1.07^{+0.06}_{-0.03}$ & $2.71^{+1.01}_{-0.62}$ & $3.69^{+0.24}_{-0.11}$ 
& $2.24^{+0.19}_{-0.15}$ & $19.5^{+0.8}_{-0.8}$ \\
\noalign{\smallskip}
\hline
\noalign{\smallskip}
\multicolumn{8}{c}{\it 14/01/1997}\\
\noalign{\smallskip}
\hline
\noalign{\smallskip}
$2 - 13$ & $0.26^{+0.02}_{-0.01}$ & $2.47^{+0.67}_{-0.54}$ & 
$1.48^{+0.04}_{-0.04}$ & $1.70^{+0.07}_{-0.04}$ & $5.50^{+0.12}_{-0.16}$ 
& $2.84^{+0.17}_{-0.19}$ & $17.8^{+0.6}_{-0.5}$ \\ 
$13 - 60$& $0.28^{+0.01}_{-0.01}$ & $2.87^{+0.58}_{-0.43}$ & 
$0.94^{+0.04}_{-0.04}$ & $1.50^{+0.10}_{-0.07}$ & $3.77^{+0.15}_{-0.13}$ 
& $2.89^{+0.41}_{-0.29}$ & $19.7^{+0.8}_{-0.6}$ \\
\noalign{\smallskip}
\hline

\end{tabular}
 
\end{table}

\clearpage

\begin{table}
\vspace{-3.0cm}
\caption{\small The characteristics of the power density spectrum. (continued) 
Parameter errors and upper limits correspond to the $1 \sigma$ and $2 
\sigma$ confidence levels respectively. \label{time_fit_1}}

\tiny

\begin{tabular}{ccccccccc}
\hline
\hline
\noalign{\smallskip}
Energy (keV)&$f_{1}^{br}$(Hz)&$\alpha_{1}$&$f_{2}^{br}$(Hz)&$\alpha_{2}$
&$f_{3}^{br}$(Hz)&$\alpha_{3}$&$rms_{BLN}$($\%$)  \\
\noalign{\smallskip}  
\hline
\noalign{\smallskip}
\multicolumn{8}{c}{\it 23/01/1997}\\
\noalign{\smallskip}
\hline
\noalign{\smallskip}
$2 - 13$ & $0.23^{+0.01}_{-0.01}$ & $3.20^{+0.54}_{-0.40}$ & 
$1.37^{+0.02}_{-0.04}$ & $1.67^{+0.04}_{-0.03}$ & $5.38^{+0.09}_{-0.08}$ & 
$2.81^{+0.12}_{-0.12}$ & $18.2^{+0.6}_{-0.4}$ \\ 
$13 - 60$& $0.22^{+0.02}_{-0.01}$ & $1.91^{+0.31}_{-0.26}$ & 
$0.98^{+0.04}_{-0.02}$ & $1.39^{+0.05}_{-0.04}$ & $3.80^{+0.09}_{-0.11}$ & 
$3.40^{+0.40}_{-0.36}$ & $19.1^{+0.8}_{-0.7}$ \\
\noalign{\smallskip}
\hline
\noalign{\smallskip}
\multicolumn{8}{c}{\it 29/01/1997}\\
\noalign{\smallskip}
\hline
\noalign{\smallskip}
$2 - 13$ & $0.30^{+0.01}_{-0.01}$ & $1.74^{+0.29}_{-0.37}$ & $1.40$ & 
$1.68^{+0.06}_{-0.04}$ & $7.40^{+0.09}_{-0.11}$ & $3.14^{+0.25}_{-0.22}$ & 
$15.3^{+0.7}_{-0.6}$ \\ 
$13 - 60$& $0.34^{+0.03}_{-0.02}$ & $1.94^{+0.28}_{-0.33}$ & 
$1.23^{+0.04}_{-0.03}$ & $1.54^{+0.10}_{-0.06}$ & $5.21^{+0.16}_{-0.22}$ & 
$3.40^{+0.77}_{-0.74}$ & $18.9^{+1.0}_{-0.9}$ \\
\noalign{\smallskip}
\hline
\noalign{\smallskip}
\multicolumn{8}{c}{\it 01/02/1997}\\
\noalign{\smallskip}
\hline
\noalign{\smallskip}
$2 - 13$ & $0.25^{+0.02}_{-0.02}$ & $1.50^{+0.18}_{-0.14}$ & 
$1.31^{+0.06}_{-0.04}$ & $1.71^{+0.12}_{-0.07}$ & $6.95^{+0.10}_{-0.10}$ & 
$2.86^{+0.17}_{-0.17}$ & $15.5^{+0.6}_{-0.5}$ \\   
$13 - 60$& $0.27^{+0.02}_{-0.02}$ & $1.26^{+0.03}_{-0.04}$ & 
$1.18^{+0.03}_{-0.02}$ & $4.18^{+1.00}_{-0.80}$ & $4.45^{+0.10}_{-0.10}$ & 
$3.73^{+0.29}_{-0.31}$ & $19.2^{+0.8}_{-0.8}$ \\
\noalign{\smallskip}
\hline
\noalign{\smallskip}
\multicolumn{8}{c}{\it 09/02/1997}\\
\noalign{\smallskip}
\hline
\noalign{\smallskip}
$2 - 13$ & $0.15^{+0.01}_{-0.01}$ & $1.85^{+0.26}_{-0.20}$ & 
$1.55^{+0.07}_{-0.06}$ & $1.61^{+0.03}_{-0.03}$ & $4.50^{+0.04}_{-0.04}$ & 
$2.71^{+0.06}_{-0.08}$ & $19.6^{+0.5}_{-0.4}$ \\   
$13 - 60$& $0.17^{+0.01}_{-0.01}$ & $1.32^{+0.07}_{-0.05}$ & 
$1.43^{+0.13}_{-0.06}$ & $2.31^{+0.23}_{-0.27}$ & $3.37^{+0.19}_{-0.20}$ & 
$2.23^{+0.18}_{-0.19}$ & $18.0^{+0.8}_{-0.7}$ \\
\noalign{\smallskip}
\hline
\noalign{\smallskip}
\multicolumn{8}{c}{\it 22/02/1997}\\
\noalign{\smallskip}
\hline
\noalign{\smallskip}
$2 - 13$ & $0.20^{+0.03}_{-0.03}$ & $1.17^{+0.29}_{-0.07}$ & 
$1.62^{+0.09}_{-0.09}$ & $1.76^{+0.46}_{-0.22}$ & $5.74^{+0.08}_{-0.08}$ & 
$3.05^{+0.21}_{-0.19}$ & $18.0^{+0.6}_{-0.7}$ \\   
$13 - 60$& $0.25^{+0.04}_{-0.05}$ & $1.26^{+0.05}_{-0.04}$ & 
$1.03^{+0.05}_{-0.06}$ & $5.05^{+0.95}_{-1.05}$ & $3.84^{+0.20}_{-0.12}$ & 
$2.94^{+0.37}_{-0.15}$ & $18.0^{+0.7}_{-0.8}$ \\
\noalign{\smallskip}
\hline
\noalign{\smallskip}
\multicolumn{8}{c}{\it 05/03/1997}\\
\noalign{\smallskip}
\hline
\noalign{\smallskip}
$2 - 13$ & $0.23^{+0.01}_{-0.02}$ & $1.98^{+0.58}_{-0.41}$ & 
$1.65^{+0.06}_{-0.11}$ & $1.69^{+0.10}_{-0.08}$ & $6.23^{+0.10}_{-0.17}$ & 
$3.02^{+0.24}_{-0.25}$ & $17.1^{+0.5}_{-0.4}$ \\   
$13 - 60$& $0.25^{+0.02}_{-0.02}$ & $1.24^{+0.81}_{-0.05}$ & 
$0.95^{+0.08}_{-0.07}$ & $2.14^{+1.50}_{-0.75}$ & $4.75^{+0.11}_{-0.20}$ & 
$3.58^{+1.16}_{-0.63}$ & $18.2^{+0.8}_{-0.7}$ \\
\noalign{\smallskip}
\hline
\noalign{\smallskip}
\multicolumn{8}{c}{\it 10/03/1997}\\
\noalign{\smallskip}
\hline
\noalign{\smallskip}
$2 - 13$ & $0.17^{+0.02}_{-0.04}$ & $2.06^{+0.50}_{-0.42}$ & 
$1.60$ & $1.63^{+0.04}_{-0.04}$ & $4.35^{+0.07}_{-0.07}$ & 
$2.65^{+0.10}_{-0.09}$ & $20.2^{+0.7}_{-0.5}$ \\   
$13 - 60$& $0.16^{+0.02}_{-0.02}$ & $1.26^{+0.05}_{-0.04}$ & 
$-$ & $-$ & $2.47^{+0.07}_{-0.05}$ & $2.15^{+0.13}_{-0.09}$ & 
$18.5^{+0.8}_{-0.6}$ \\
\noalign{\smallskip}
\hline
\noalign{\smallskip}
\multicolumn{8}{c}{\it 17/03/1997}\\
\noalign{\smallskip}
\hline
\noalign{\smallskip}
$2 - 13$ & $0.24^{+0.01}_{-0.01}$ & $1.90^{+0.33}_{-0.36}$ & 
$1.66^{+0.06}_{-0.07}$ & $1.60^{+0.04}_{-0.03}$ & $6.35^{+0.10}_{-0.10}$ & 
$3.35^{+0.25}_{-0.51}$ & $17.3^{+0.6}_{-0.7}$ \\   
$13 - 60$& $0.26^{+0.01}_{-0.01}$ & $1.51^{+0.41}_{-0.06}$ & 
$0.98^{+0.07}_{-0.06}$ & $1.39^{+1.80}_{-0.08}$ & $4.32^{+0.06}_{-0.12}$ & 
$3.91^{+0.50}_{-0.64}$ & $18.1^{+0.8}_{-0.9}$ \\
\noalign{\smallskip}
\hline
\noalign{\smallskip}
\multicolumn{8}{c}{\it 26/03/1997}\\
\noalign{\smallskip}
\hline
\noalign{\smallskip}
$2 - 13$ & $0.34^{+0.03}_{-0.02}$ & $2.51^{+0.56}_{-0.62}$ & 
$1.74^{+0.07}_{-0.09}$ & $1.68^{+0.06}_{-0.05}$ & $7.28^{+0.11}_{-0.20}$ & 
$3.91^{+0.53}_{-0.47}$ & $16.4^{+0.6}_{-0.6}$ \\   
$13 - 60$& $0.37^{+0.03}_{-0.02}$ & $2.12^{+0.73}_{-0.58}$ & 
$1.18^{+0.16}_{-0.10}$ & $1.27^{+0.08}_{-0.09}$ & $4.79^{+0.09}_{-0.21}$ & 
$4.00^{+1.02}_{-0.97}$ & $18.0^{+0.9}_{-0.9}$ \\
\noalign{\smallskip}
\hline
\noalign{\smallskip}
\multicolumn{8}{c}{\it 27/03/1997}\\
\noalign{\smallskip}
\hline
\noalign{\smallskip}
$2 - 13$ & $0.34^{+0.02}_{-0.02}$ & $3.78^{+1.16}_{-0.86}$ & 
$1.80$ & $2.34^{+0.39}_{-0.34}$ & $7.25^{+0.15}_{-0.26}$ & 
$2.66^{+0.27}_{-0.27}$ & $15.9^{+0.6}_{-0.6}$ \\   
$13 - 60$& $0.33^{+0.03}_{-0.02}$ & $1.91^{+0.71}_{-0.41}$ & 
$1.34^{+0.08}_{-0.08}$ & $4.20^{+0.51}_{-1.78}$ & $4.87^{+0.23}_{-0.18}$ & 
$2.38^{+0.29}_{-0.18}$ & $18.0^{+0.9}_{-0.8}$ \\
\noalign{\smallskip}
\hline
\noalign{\smallskip}
\multicolumn{8}{c}{\it 02/04/1997$^{*}$}\\
\noalign{\smallskip}
\hline
\noalign{\smallskip}
$2 - 13$ & $1.01^{+0.03}_{-0.03}$ & $1.67^{+0.13}_{-0.06}$ & 
$-$ & $-$ & $10.35^{+0.28}_{-0.23}$ & $3.52^{+0.38}_{-0.38}$ & 
$11.4^{+0.5}_{-0.5}$ \\   
$13 - 60$& $0.81^{+0.04}_{-0.01}$ & $1.47^{+0.05}_{-0.03}$ & 
$-$ & $-$ & $6.78^{+0.56}_{-0.80}$ & $3.55^{+0.60}_{-0.43}$ & 
$16.7^{+0.8}_{-0.9}$ \\ 
\noalign{\smallskip}
\hline
\noalign{\smallskip}
\multicolumn{8}{c}{\it 10/04/1997$^{*}$}\\
\noalign{\smallskip}
\hline
\noalign{\smallskip}
$2 - 13$ & $0.95^{+0.08}_{-0.05}$ & $1.66^{+0.13}_{-0.10}$ & 
$-$ & $-$ & $10.98^{+0.43}_{-0.43}$ & $4.73^{+0.56}_{-1.22}$ & 
$10.9^{+0.6}_{-0.7}$ \\   
$13 - 60$& $0.67^{+0.04}_{-0.04}$ & $1.41^{+0.08}_{-0.07}$ & 
$-$ & $-$ & $5.55^{+0.18}_{-0.18}$ & $3.85^{+0.75}_{-0.50}$ & 
$18.4^{+0.9}_{-1.0}$ \\ 
\noalign{\smallskip}
\hline
\noalign{\smallskip}
\multicolumn{8}{c}{\it 19/04/1997}\\
\noalign{\smallskip}
\hline
\noalign{\smallskip}
$2 - 13$ & $0.17^{+0.02}_{-0.01}$ & $2.21^{+0.54}_{-0.44}$ & 
$1.31^{+0.12}_{-0.05}$ & $1.58^{+0.06}_{-0.04}$ & $5.10^{+0.11}_{-0.10}$ & 
$2.82^{+0.18}_{-0.17}$ & $17.8^{+0.6}_{-0.7}$ \\   
$13 - 60$& $0.19^{+0.01}_{-0.03}$ & $1.63^{+0.27}_{-0.25}$ & 
$1.28^{+0.02}_{-0.20}$ & $3.39^{+0.59}_{-0.94}$ & $4.50^{+0.20}_{-0.14}$ & 
$2.00^{+0.05}_{-0.02}$ & $17.2^{+0.8}_{-0.9}$ \\
\noalign{\smallskip}
\hline
\noalign{\smallskip}
\multicolumn{8}{c}{\it 23/04/1997}\\
\noalign{\smallskip}
\hline
\noalign{\smallskip}
$2 - 13$ & $0.22^{+0.02}_{-0.01}$ & $2.96^{+0.79}_{-0.64}$ & 
$1.42^{+0.05}_{-0.05}$ & $1.77^{+0.13}_{-0.07}$ & $4.97^{+0.16}_{-0.19}$ & 
$2.46^{+0.18}_{-0.19}$ & $18.0^{+0.5}_{-0.6}$ \\   
$13 - 60$& $0.20^{+0.01}_{-0.02}$ & $1.62^{+0.28}_{-0.06}$ & 
$0.87^{+0.03}_{-0.03}$ & $1.47^{+0.26}_{-0.07}$ & $4.00^{+0.15}_{-0.17}$ & 
$3.14^{+0.65}_{-0.48}$ & $18.3^{+0.7}_{-0.7}$ \\
\noalign{\smallskip}
\hline
\noalign{\smallskip}
\multicolumn{8}{c}{\it 25/04/1997}\\
\noalign{\smallskip}
\hline
\noalign{\smallskip}
$2 - 13$ & $0.34^{+0.01}_{-0.01}$ & $1.56^{+0.11}_{-0.08}$ & 
$1.63^{+0.03}_{-0.02}$ & $3.18^{+0.26}_{-0.24}$ & $8.92^{+0.07}_{-0.07}$ & 
$2.81^{+0.12}_{-0.11}$ & $13.5^{+0.5}_{-0.5}$ \\   
$13 - 60$& $0.34^{+0.01}_{-0.01}$ & $1.37^{+0.06}_{-0.04}$ & 
$1.48^{+0.06}_{-0.03}$ & $4.23^{+0.54}_{-0.58}$ & $5.80^{+0.16}_{-0.19}$ & 
$3.39^{+0.41}_{-0.42}$ & $18.8^{+0.8}_{-0.7}$ \\
\noalign{\smallskip}
\hline

\end{tabular}
\begin{list}{}{}
\item[$^{*}$] -- for these observations the additional power law 
noise component was included to the fit to describe the low frequency 
part of the power density spectrum.
\end{list} 
\end{table}

\clearpage

\begin{table}
\vspace{-3.0cm}
\caption{\small The characteristics of the power density spectrum 
not included to the Tables \ref{time_fit}, \ref{time_fit_1}. 
\label{time_con}}

\tiny

\begin{tabular}{ccccccc}
\hline
\hline
\noalign{\smallskip}
$f_{QPO}$(Hz)&FWHM (Hz)&$rms^{QPO}_{1}$($\%$)&$rms^{QPO}_{1/2}$($\%$)&
$rms^{QPO}_{2}$($\%$)&$rms^{QPO}_{3}$($\%$)&$\chi^{2}(d.o.f)$ \\
\noalign{\smallskip}  
\hline
\noalign{\smallskip}
\multicolumn{7}{c}{\it 23/10/1996}\\
\noalign{\smallskip}
\hline
\noalign{\smallskip}
$5.02^{+0.01}_{-0.02}$ & $0.71^{+0.03}_{-0.04}$ & $5.9^{+0.4}_{-0.5}$ & 
$<2.5$ & $<1.5$ & $<1.0$ & $381.0(205)$ \\
$5.14^{+0.07}_{-0.03}$ & $0.66^{+0.09}_{-0.07}$ & $7.0^{+0.6}_{-0.5}$ & 
$3.7^{+0.8}_{-0.7}$ & $<2.2$ & $<1.7$ & $256.0(144)$\\ 
$5.07^{+0.01}_{-0.01}$ & $0.66^{+0.04}_{-0.03}$ & $5.4^{+0.5}_{-0.5}$ & 
$1.9^{+0.6}_{-0.6}$ & $<2.0$ & $<1.2$ & $371.(205)$\\
$5.15^{+0.10}_{-0.10}$ & $0.83^{+0.10}_{-0.10}$ & $7.0^{+0.6}_{-0.6}$ & 
$3.5^{+0.7}_{-0.8}$ & $2.3^{+0.8}_{-0.7}$ & $<1.8$ & $268.5(144)$\\
$5.12^{+0.02}_{-0.01}$ & $0.77^{+0.05}_{-0.04}$ & $5.4^{+0.5}_{-0.5}$ & 
$2.3^{+0.5}_{-0.5}$ & $<2.1$ & $<1.0$ & $322.7(205)$ \\ 
$5.20^{+0.05}_{-0.10}$ & $0.79^{+0.10}_{-0.10}$ & $6.8^{+0.5}_{-0.7}$ & 
$3.8^{+0.9}_{-0.7}$ & $<2.5$ & $<2.0$ & $234.5(144)$\\
$4.45^{+0.01}_{-0.01}$ & $0.70^{+0.03}_{-0.04}$ & $7.2^{+0.6}_{-0.5}$ & 
$2.9^{+0.4}_{-0.4}$ & $<2.0$ & $<1.3$ & $383.1(205)$ \\
$4.51^{+0.03}_{-0.02}$ & $0.64^{+0.07}_{-0.06}$ & $8.6^{+0.5}_{-0.5}$ & 
$4.0^{+0.9}_{-0.6}$ & $<2.0$ & $<2.5$ & $237.8(144)$ \\ 
\noalign{\smallskip}
\hline
\noalign{\smallskip}
\multicolumn{7}{c}{\it 29/10/1996}\\
\noalign{\smallskip}
\hline
\noalign{\smallskip}
$3.22^{+0.01}_{-0.01}$ & $0.65^{+0.01}_{-0.02}$ & $12.2^{+0.2}_{-0.3}$ & 
$<0.5$ & $3.5^{+0.3}_{-0.3}$ & $0.8^{+0.4}_{-0.4}$ & $251.7(205)$ \\ 
$3.23^{+0.01}_{-0.01}$ & $0.69^{+0.01}_{-0.02}$ & $16.2^{+0.3}_{-0.4}$ & 
$<1.2$ & $2.7^{+0.7}_{-0.6}$ & $2.4^{+0.7}_{-0.8}$ & $207.1(144)$\\
\noalign{\smallskip}
\hline
\noalign{\smallskip}
\multicolumn{7}{c}{\it 07/11/1996}\\
\noalign{\smallskip}
\hline
\noalign{\smallskip}
$4.81^{+0.01}_{-0.01}$ & $0.57^{+0.02}_{-0.03}$ & $6.4^{+0.5}_{-0.6}$ & 
$1.5^{+0.5}_{-0.6}$ & $<2.3$ & $<1.9$ & $380.8(205)$ \\ 
$4.88^{+0.02}_{-0.05}$ & $0.51^{+0.05}_{-0.05}$ & $7.5^{+0.6}_{-0.5}$ & 
$2.2^{+0.6}_{-0.5}$ & $<3.0$ & $<1.3$ & $238.3(144)$\\
$5.08^{+0.01}_{-0.01}$ & $0.68^{+0.02}_{-0.02}$ & $5.9^{+0.5}_{-0.5}$ & 
$1.5^{+0.5}_{-0.5}$ & $<2.0$ & $<1.3$ & $430.9(205)$ \\ 
$5.15^{+0.05}_{-0.08}$ & $0.71^{+0.05}_{-0.06}$ & $7.4^{+0.6}_{-0.6}$ & 
$3.5^{+0.8}_{-0.8}$ & $2.5^{+0.4}_{-0.5}$ & $<1.6$ & $321.3(144)$\\
$7.83^{+0.04}_{-0.04}$ & $0.80^{+0.10}_{-0.08}$ & $2.6^{+0.5}_{-0.6}$ & 
$-$ & $-$ & $-$ & $272.5(178)$\\
$7.83^{+0.04}_{-0.04}$ & $0.98^{+0.14}_{-0.14}$ & $7.4^{+0.8}_{-0.9}$ & 
$-$ & $-$ & $-$ & $137.7(93)$ \\ 
$4.20^{+0.03}_{-0.03}$ & $1.22^{+0.10}_{-0.10}$ & $8.7^{+0.6}_{-0.6}$ & 
$3.5^{+0.8}_{-0.9}$ & $<2.3$ & $<1.8$ & $290.7(205)$\\
$4.25^{+0.05}_{-0.05}$ & $1.20^{+0.20}_{-0.19}$ & $11.1^{+0.7}_{-0.7}$ & 
$4.8^{+0.8}_{-0.7}$ & $-$ & $-$ & $107.6(88)$ \\ 
\noalign{\smallskip}
\hline
\noalign{\smallskip}
\multicolumn{7}{c}{\it 14/11/1996}\\
\noalign{\smallskip}
\hline
\noalign{\smallskip}
$4.25^{+0.02}_{-0.02}$ & $0.68^{+0.03}_{-0.03}$ & $8.0^{+0.3}_{-0.3}$ & 
$2.5^{+0.5}_{-0.5}$ & $<1.0$ & $<1.0$ & $304.9(205)$ \\ 
$4.28^{+0.01}_{-0.03}$ & $0.63^{+0.06}_{-0.05}$ & $8.6^{+0.5}_{-0.5}$ 
& $3.1^{+0.8}_{-0.9}$ & $2.2^{+0.8}_{-0.8}$ & $<2.0$ & $240.2(144)$\\
\noalign{\smallskip}
\hline
\noalign{\smallskip}
\multicolumn{7}{c}{\it 28/11/1996}\\
\noalign{\smallskip} 
\hline
$4.63^{+0.03}_{-0.01}$ & $0.94^{+0.02}_{-0.03}$ & $7.9^{+0.2}_{-0.2}$ & 
$2.4^{+0.3}_{-0.4}$ & $<1.5$ & $<1.0$ & $356.1(205)$ \\ 
$4.65^{+0.02}_{-0.02}$ & $1.03^{+0.05}_{-0.06}$ & $9.4^{+0.3}_{-0.3}$ & 
$3.2^{+0.7}_{-0.7}$ & $<1.6$ & $<2.0$ & $224.7(144)$\\
\noalign{\smallskip}
\hline
\noalign{\smallskip}
\multicolumn{7}{c}{\it 04/12/1996}\\
\noalign{\smallskip}
\hline
\noalign{\smallskip}
$2.81^{+0.01}_{-0.01}$ & $0.41^{+0.01}_{-0.01}$ & $11.1^{+0.4}_{-0.4}$ & 
$2.2^{+0.3}_{-0.4}$ & $5.0^{+0.2}_{-0.3}$ & $1.5^{+0.3}_{-0.3}$ &
$418.2(208)$ \\ 
$2.81^{+0.01}_{-0.01}$ & $0.42^{+0.02}_{-0.02}$ & $13.5^{+0.4}_{-0.5}$ 
& $2.3^{+0.7}_{-0.7}$ & $3.4^{+0.5}_{-0.4}$ & $<1.6$ & $164.1(144)$\\
\noalign{\smallskip}
\hline
\noalign{\smallskip}
\multicolumn{7}{c}{\it 11/12/1996}\\
\noalign{\smallskip} 
\hline
$2.77^{+0.01}_{-0.01}$ & $0.81^{+0.02}_{-0.02}$ & $13.0^{+0.1}_{-0.3}$ & 
$<2.2$ & $6.3^{+0.05}_{-0.05}$ & $2.3^{+0.4}_{-0.3}$ & 290.6(205)\\
$2.78^{+0.01}_{-0.01}$ & $0.80^{+0.04}_{-0.03}$ & $14.9^{+0.5}_{-0.3}$ & 
$2.6^{+0.5}_{-0.4}$ & $4.00^{+0.5}_{-0.4}$ & $<1.4$ & $176.6(144)$ \\
\noalign{\smallskip} 
\hline
\noalign{\smallskip}
\multicolumn{7}{c}{\it 19/12/1996}\\
\noalign{\smallskip} 
\hline
\noalign{\smallskip}
$3.11^{+0.01}_{-0.01}$ & $0.76^{+0.03}_{-0.02}$ & $12.1^{+0.3}_{-0.2}$ & 
$3.7^{+0.5}_{-0.5}$ & $5.5^{+0.2}_{-0.3}$ & $2.1^{+0.5}_{-0.4}$ & 235.5(205)\\
$3.12^{+0.01}_{-0.01}$ & $0.73^{+0.03}_{-0.03}$ & $13.3^{+0.5}_{-0.3}$ & 
$3.2^{+0.9}_{-1.0}$ & $3.1^{+0.5}_{-0.5}$ & $<1.1$ & $147.9(144)$\\
\noalign{\smallskip}
\hline
\noalign{\smallskip}
\multicolumn{7}{c}{\it 24/12/1996}\\
\noalign{\smallskip} 
\hline
\noalign{\smallskip}
$3.90^{+0.01}_{-0.01}$ & $0.79^{+0.03}_{-0.03}$ & $10.2^{+0.3}_{-0.3}$ &
$2.5^{+0.5}_{-0.5}$ & $2.7^{+0.4}_{-0.4}$ & $1.1^{+0.4}_{-0.4}$ & 362.3(205) \\
$3.91^{+0.02}_{-0.01}$ & $0.74^{+0.05}_{-0.04}$ & $10.3^{+0.4}_{-0.4}$ & 
$2.6^{+0.8}_{-0.8}$ & $<1.0$ & $<1.0$ & $209.5(144)$\\
\noalign{\smallskip}
\hline
\noalign{\smallskip}
\multicolumn{7}{c}{\it 31/12/1996}\\
\noalign{\smallskip}
\hline
\noalign{\smallskip}
$2.80^{+0.01}_{-0.01}$ & $0.61^{+0.02}_{-0.03}$ & $11.8^{+0.2}_{-0.4}$ & 
$2.9^{+0.6}_{-0.5}$ & $6.2^{+0.9}_{-1.0}$ & $2.1^{+0.4}_{-0.4}$ & 
229.4(205) \\ 
$2.81^{+0.01}_{-0.01}$ & $0.50^{+0.03}_{-0.02}$ & $11.9^{+0.4}_{-0.4}$ & 
$2.7^{+0.8}_{-0.8}$ & $2.9^{+0.5}_{-0.5}$ & $<1.0$ & $132.8(144)$\\ 
\noalign{\smallskip}
\hline
\noalign{\smallskip}
\multicolumn{7}{c}{\it 07/01/1997}\\
\noalign{\smallskip}
\hline
\noalign{\smallskip}
$2.91^{+0.01}_{-0.01}$ & $0.67^{+0.02}_{-0.02}$ & $12.1^{+0.6}_{-0.3}$ & 
$2.3^{+0.6}_{-0.6}$ & $5.9^{+0.2}_{-0.2}$ & $2.1^{+0.3}_{-0.3}$ & 
298.0(205) \\
$2.91^{+0.01}_{-0.01}$ & $0.57^{+0.03}_{-0.02}$ & $12.0^{+0.4}_{-0.4}$ & 
$2.7^{+0.3}_{-0.3}$ & $3.1^{+0.5}_{-0.5}$ & $<1.0$ & $170.6(144)$\\
\noalign{\smallskip}
\hline
\noalign{\smallskip}
\multicolumn{7}{c}{\it 14/01/1997}\\
\noalign{\smallskip}
\hline
\noalign{\smallskip}
$2.92^{+0.01}_{-0.01}$ & $0.63^{+0.03}_{-0.02}$ & $11.6^{+0.3}_{-0.3}$ & 
$2.1^{+0.8}_{-0.8}$ & $5.5^{+0.3}_{-0.3}$ & $1.9^{+0.3}_{-0.3}$ & 
220.5(205) \\
$2.92^{+0.01}_{-0.01}$ & $0.54^{+0.04}_{-0.03}$ & $11.1^{+0.4}_{-0.4}$ & 
$2.1^{+0.6}_{-0.6}$ & $2.5^{+0.3}_{-0.3}$ & $<1.0$ & $163.7(144)$\\
\noalign{\smallskip}
\hline
\noalign{\smallskip}
\multicolumn{7}{c}{\it 23/01/1997}\\
\noalign{\smallskip} 
\hline
$2.80^{+0.01}_{-0.01}$ & $0.58^{+0.01}_{-0.02}$ & $12.0^{+0.3}_{-0.3}$ & 
$2.2^{+0.6}_{-0.5}$ & $4.5^{+0.4}_{-0.4}$ & $<2.2$ & $316.2(205)$ \\ 
$2.80^{+0.01}_{-0.01}$ & $0.48^{+0.03}_{-0.02}$ & $11.4^{+0.3}_{-0.3}$ & 
$2.3^{+0.5}_{-0.5}$ & $3.5^{+0.5}_{-0.6}$ & $<1.0$ & $157.2(144)$\\
\noalign{\smallskip}
\hline

\end{tabular}
 
\end{table}

\begin{table}
\vspace{-3.0cm}
\caption{\small The characteristics of the power density spectrum not 
included to the Tables \ref{time_fit}, \ref{time_fit_1}. (continued) 
\label{time_con_1}}

\tiny

\begin{tabular}{ccccccc}
\hline
\hline
\noalign{\smallskip}
$f_{QPO}$(Hz)&FWHM (Hz)&$rms^{QPO}_{1}$($\%$)&$rms^{QPO}_{1/2}$($\%$)&
$rms^{QPO}_{2}$($\%$)&$rms^{QPO}_{3}$($\%$)&$\chi^{2}(d.o.f)$ \\
\noalign{\smallskip}
\multicolumn{7}{c}{\it 29/01/1997}\\
\noalign{\smallskip} 
\hline
$3.64^{+0.02}_{-0.01}$ & $0.99^{+0.03}_{-0.02}$ & $11.4^{+0.2}_{-0.3}$ & 
$3.2^{+0.4}_{-0.4}$ & $3.6^{+0.8}_{-0.9}$ & $1.6^{+0.5}_{-0.6}$ & 
$296.0(202)$ \\ 
$3.67^{+0.01}_{-0.02}$ & $0.90^{+0.04}_{-0.05}$ & $11.0^{+0.3}_{-0.3}$ & 
$1.9^{+0.6}_{-0.6}$ & $1.8^{+0.6}_{-0.6}$ & $<1.5$ & $200.7(145)$\\
\noalign{\smallskip}
\hline
\noalign{\smallskip}
\multicolumn{7}{c}{\it 01/02/1997}\\
\noalign{\smallskip} 
\hline
$3.56^{+0.01}_{-0.01}$ & $0.96^{+0.03}_{-0.03}$ & $11.4^{+0.2}_{-0.2}$ & 
$3.7^{+0.4}_{-0.4}$ & $3.8^{+0.4}_{-0.4}$ & $1.3^{+0.4}_{-0.3}$ & 
$342.8(205)$ \\ 
$3.55^{+0.01}_{-0.01}$ & $0.72^{+0.04}_{-0.05}$ & $10.0^{+0.3}_{-0.3}$ & 
$2.9^{+0.6}_{-0.5}$ & $3.0^{+0.6}_{-0.6}$ & $<2.0$ & $195.8(144)$\\
\noalign{\smallskip}
\hline
\noalign{\smallskip}
\multicolumn{7}{c}{\it 09/02/1997}\\
\noalign{\smallskip} 
\hline
$2.26^{+0.01}_{-0.01}$ & $0.43^{+0.01}_{-0.01}$ & $12.4^{+0.3}_{-0.3}$ & 
$2.6^{+0.4}_{-0.3}$ & $5.4^{+0.2}_{-0.2}$ & $2.0^{+0.6}_{-0.5}$ &
$251.5(205)$ \\ 
$2.26^{+0.01}_{-0.01}$ & $0.45^{+0.02}_{-0.02}$ & $12.4^{+0.3}_{-0.3}$ & 
$1.6^{+0.7}_{-0.6}$ & $4.0^{+0.4}_{-0.4}$ & $1.5^{+0.5}_{-0.5}$ & 
$158.5(144)$\\
\noalign{\smallskip}
\hline
\noalign{\smallskip}
\multicolumn{7}{c}{\it 22/02/1997}\\
\noalign{\smallskip} 
\hline
$2.98^{+0.01}_{-0.01}$ & $0.61^{+0.02}_{-0.02}$ & $11.3^{+0.3}_{-0.3}$ & 
$3.2^{+0.5}_{-0.4}$ & $4.0^{+0.6}_{-0.5}$ & $<0.8$ & $232.4(205)$ \\ 
$2.98^{+0.01}_{-0.01}$ & $0.51^{+0.03}_{-0.03}$ & $10.3^{+0.3}_{-0.3}$ & 
$3.3^{+0.6}_{-0.6}$ & $2.9^{+0.6}_{-0.6}$ & $<1.3$ & $175.4(144)$\\
\noalign{\smallskip}
\hline
\noalign{\smallskip}
\multicolumn{7}{c}{\it 05/03/1997}\\
\noalign{\smallskip} 
\hline
$3.24^{+0.01}_{-0.01}$ & $0.64^{+0.04}_{-0.03}$ & $11.1^{+0.4}_{-0.4}$ & 
$2.9^{+0.8}_{-0.7}$ & $3.8^{+0.7}_{-0.7}$ & $<1.1$ & $270.1(205)$ \\ 
$3.25^{+0.01}_{-0.01}$ & $0.61^{+0.05}_{-0.04}$ & $11.0^{+0.4}_{-0.3}$ & 
$3.4^{+0.8}_{-0.8}$ & $1.8^{+0.6}_{-0.6}$ & $<2.2$ & $123.1(144)$\\
\noalign{\smallskip}
\hline
\noalign{\smallskip}
\multicolumn{7}{c}{\it 10/03/1997}\\
\noalign{\smallskip} 
\hline
$2.21^{+0.01}_{-0.01}$ & $0.35^{+0.02}_{-0.02}$ & $12.2^{+0.4}_{-0.4}$ & 
$2.4^{+0.5}_{-0.6}$ & $5.5^{+0.4}_{-0.4}$ & $<1.0$ & $257.4(205)$ \\ 
$2.21^{+0.01}_{-0.01}$ & $0.33^{+0.01}_{-0.02}$ & $11.1^{+0.4}_{-0.4}$ & 
$2.4^{+0.5}_{-0.5}$ & $3.7^{+0.4}_{-0.3}$ & $<1.2$ & $139.7(144)$\\
\noalign{\smallskip}
\hline
\noalign{\smallskip}
\multicolumn{7}{c}{\it 17/03/1997}\\
\noalign{\smallskip} 
\hline
$3.21^{+0.01}_{-0.01}$ & $0.64^{+0.03}_{-0.02}$ & $11.3^{+0.4}_{-0.4}$ & 
$2.8^{+0.5}_{-0.5}$ & $4.0^{+0.4}_{-0.5}$ & $<1.5$ & $256.3(205)$ \\ 
$3.21^{+0.01}_{-0.01}$ & $0.58^{+0.03}_{-0.03}$ & $10.8^{+0.4}_{-0.3}$ & 
$2.7^{+0.4}_{-0.5}$ & $3.3^{+0.5}_{-0.5}$ & $<1.4$ & $140.3(144)$\\
\noalign{\smallskip}
\hline
\noalign{\smallskip}
\multicolumn{7}{c}{\it 26/03/1997}\\
\noalign{\smallskip} 
\hline
$3.52^{+0.01}_{-0.01}$ & $0.71^{+0.05}_{-0.03}$ & $10.6^{+0.4}_{-0.5}$ & 
$1.7^{+0.8}_{-0.7}$ & $3.1^{+0.6}_{-0.6}$ & $1.8^{+0.6}_{-0.6}$ & 
$214.0(205)$ \\ 
$3.52^{+0.01}_{-0.01}$ & $0.70^{+0.06}_{-0.07}$ & $10.3^{+0.5}_{-0.5}$ & 
$2.3^{+1.1}_{-0.9}$ & $2.3^{+0.5}_{-0.6}$ & $<1.5$ & $145.8(144)$\\
\noalign{\smallskip}
\hline
\noalign{\smallskip}
\multicolumn{7}{c}{\it 27/03/1997}\\
\noalign{\smallskip} 
\hline
$3.65^{+0.01}_{-0.01}$ & $0.84^{+0.05}_{-0.04}$ & $10.9^{+0.5}_{-0.5}$ & 
$<1.5$& $3.3^{+0.8}_{-0.8}$ & $<1.5$ & $206.0(205)$ \\ 
$3.64^{+0.01}_{-0.01}$ & $0.77^{+0.07}_{-0.07}$ & $10.7^{+0.3}_{-0.5}$ & 
$3.0^{+0.6}_{-0.7}$ & $2.5^{+0.5}_{-0.6}$ & $<1.3$ & $129.2(144)$\\
\noalign{\smallskip}
\hline
\noalign{\smallskip}
\multicolumn{7}{c}{\it 02/04/1997}\\
\noalign{\smallskip} 
\hline
$4.90^{+0.01}_{-0.03}$ & $1.47^{+0.06}_{-0.05}$ & $9.2^{+0.3}_{-0.3}$ & 
$3.4^{+0.4}_{-0.4}$ & $2.1^{+0.7}_{-0.7}$ & $1.3^{+0.6}_{-0.6}$ & 
$311.2(206)$ \\ 
$4.86^{+0.03}_{-0.02}$ & $1.60^{+0.12}_{-0.05}$ & $10.4^{+0.5}_{-0.5}$ & 
$2.4^{+0.8}_{-0.8}$ & $<2.1$ & $<1.5$ & $168.8(144)$\\
\noalign{\smallskip}
\hline
\noalign{\smallskip}
\multicolumn{7}{c}{\it 10/04/1997$^{*}$}\\
\noalign{\smallskip} 
\hline
$4.44^{+0.01}_{-0.01}$ & $1.29^{+0.05}_{-0.05}$ & $10.6^{+0.4}_{-0.4}$ & 
$4.1^{+0.4}_{-0.4}$ & $4.1^{+0.6}_{-0.6}$ & $<2.4$ & $283.9(206)$ \\
$4.45^{+0.03}_{-0.03}$ & $0.82^{+0.12}_{-0.10}$ & $8.3^{+0.5}_{-0.6}$ & 
$<3.0$ & $2.9^{+0.9}_{-0.9}$ & $<3.0$ & $179.7(148)$ \\ 
\noalign{\smallskip}
\hline
\noalign{\smallskip}
\multicolumn{7}{c}{\it 19/04/1997}\\
\noalign{\smallskip} 
\hline
$2.53^{+0.01}_{-0.01}$ & $0.61^{+0.03}_{-0.03}$ & $12.8^{+0.4}_{-0.4}$ & 
$2.2^{+0.8}_{-0.7}$ & $5.6^{+0.6}_{-0.5}$ & $<0.7$ & $209.8(205)$ \\ 
$2.55^{+0.01}_{-0.02}$ & $0.77^{+0.04}_{-0.04}$ & $13.4^{+0.4}_{-0.3}$ & 
$1.2^{+0.6}_{-0.5}$ & $3.4^{+0.6}_{-0.7}$ & $<1.0$ & $176.5(144)$\\
\noalign{\smallskip}
\hline
\noalign{\smallskip}
\multicolumn{7}{c}{\it 23/04/1997}\\
\noalign{\smallskip} 
\hline
$2.72^{+0.01}_{-0.01}$ & $0.59^{+0.03}_{-0.03}$ & $12.3^{+0.4}_{-0.4}$ & 
$1.7^{+0.8}_{-0.8}$ & $5.4^{+0.6}_{-0.6}$ & $<1.3$ & $233.0(205)$ \\ 
$2.71^{+0.01}_{-0.01}$ & $0.58^{+0.03}_{-0.03}$ & $12.5^{+0.5}_{-0.4}$ & 
$2.8^{+0.7}_{-0.7}$ & $3.7^{+0.3}_{-0.4}$ & $<1.7$ & $120.5(144)$\\
\noalign{\smallskip}
\hline
\noalign{\smallskip}
\multicolumn{7}{c}{\it 25/04/1997}\\
\noalign{\smallskip} 
\hline
$4.27^{+0.01}_{-0.01}$ & $1.14^{+0.03}_{-0.03}$ & $9.9^{+0.2}_{-0.2}$ & 
$3.1^{+0.4}_{-0.4}$ & $5.9^{+0.5}_{-0.4}$ & $1.3^{+0.4}_{-0.3}$ & 
$458.0(205)$ \\ 
$4.28^{+0.02}_{-0.01}$ & $1.05^{+0.06}_{-0.06}$ & $9.9^{+0.3}_{-0.2}$ & 
$2.8^{+0.6}_{-0.6}$ & $2.5^{+0.5}_{-0.4}$ & $1.7^{+0.8}_{-0.9}$ & 
$232.5(144)$\\
\noalign{\smallskip}
\hline

\end{tabular}
 
\end{table}

\clearpage

\clearpage

\clearpage

\clearpage

\clearpage

\clearpage

\clearpage

\clearpage

\clearpage

\clearpage

\clearpage

\clearpage

\clearpage

\clearpage

\end{document}